\documentclass[12pt]{article}
\usepackage{mathptmx,helvet,courier,graphicx,
makeidx,multicol,footmisc,comment}
\usepackage{latexsym,amssymb,amsfonts,graphicx}
\usepackage{amsmath}
\usepackage{amssymb}
\usepackage{amsthm}
\usepackage{latexsym}
\usepackage{color}
\usepackage{enumerate}
\usepackage{graphicx}

\usepackage{color}

\newcommand{\eps}{\varepsilon}

\newcommand{\ba}{\begin{array}}
\newcommand{\ea}{\end{array}}
\newcommand{\be}{\begin{equation}}
\newcommand{\ee}{\end{equation}}
\newcommand{\bea}{\begin{eqnarray}}
\newcommand{\eea}{\end{eqnarray}}
\newcommand{\beaa}{\begin{eqnarray*}}
\newcommand{\eeaa}{\end{eqnarray*}}

\def\g{\gamma}

\def\q{\quad}

\def\pa{\partial}

\newcommand{\basa}{\begin{assumption}}
\newcommand{\easa}{\end{assumption}}

\newcommand{\bas}{\begin{assum}}
\newcommand{\eas}{\end{assum}}

\newcommand{\No}[1]{\left\|#1\right\|}     
\newcommand{\abs}[1]{\left|#1\right|}     

\newtheorem{Theorem}{Theorem}[part]
\newtheorem{Definition}{Definition}[part]
\newtheorem{Proposition}{Proposition}[part]
\newtheorem{Assumption}{Assumption}[part]

\newtheorem{Remark}{Remark}[part]

\def \ep{\hbox{ }\hfill$\Box$}
\def\reff#1{{\rm(\ref{#1})}}

\newtheorem{thm}{Theorem}[section]

\newtheorem{prop}[thm]{Proposition}
\newtheorem{rem}[thm]{Remark}

\newtheorem{assum}[thm]{Assumption}

\def\nobs{{\mbox{\scriptsize obs}\!\!\!\!\!\!\!\!\!\!\!-\!\!-}}
\def\obs{{\mbox{\scriptsize obs}}}

\begin{document}
\title{\textbf{Moral Hazard in Dynamic Risk Management}\thanks{We would like to express our gratitude to Paolo Guasoni, Yuliy Sannikov, Julio Backhoff, and participants  at: the Caltech brown bag seminar, the Bachelier seminar, the 2014 Risk and Stochastics conference at LSE,  the 2014
Arbitrage and Portfolio Optimization conference at BIRS, Banff, the 2014 Oberwolfach conference on Stochastic Analysis in Finance and Insurance,
the 2014 CRETE Conference on Research on Economic Theory \& Econometrics,
the 5th SIAM Conference on Financial Mathematics \& Engineering, and the 2014 conference on  Optimization meets general equilibrium theory, dynamic contracting and finance at University of Chile.
}
}
\author{
Jak\v sa Cvitani\'{c}\thanks{
Caltech, Humanities and Social
Sciences, M/C 228-77,
1200 E. California Blvd. Pasadena, CA 91125, USA;
E-mail: cvitanic@hss.caltech.edu.
 Research
supported in part by NSF grant DMS 10-08219.}, \ \ Dylan Possama\"i\thanks{CEREMADE, University Paris-Dauphine, place du Mar\'echal de Lattre de Tassigny, 75116 Paris, France; Email: possamai@ceremade.dauphine.fr}
  \ \ and \ \ Nizar Touzi\thanks{CMAP, Ecole Polytechnique, Route de Saclay, 91128 Palaiseau, France; Email: nizar.touzi@polytechnique.edu}}
\maketitle



\noindent\textbf{Abstract.}
We consider a contracting problem in which a principal
hires an agent to manage a risky project.
When the agent chooses  volatility components of the output process
and the principal observes the output continuously, the principal
can compute the quadratic variation of the output, but not the
individual components. This leads to moral hazard with respect to
the risk choices of the agent. We identify a family of admissible contracts for which
 the optimal agent's action is explicitly characterized, and, using the  recent theory of singular changes
of measures for It\^o processes, we study how restrictive this family is. In particular, in the special case of the
  standard Homlstrom-Milgrom model with fixed volatility,
  the family includes all possible contracts.
 We solve the  principal-agent
problem in the case of CARA preferences,
and  show that the optimal contract is linear in these  factors:
 the contractible sources of risk, including the output,
  the  quadratic variation of the output and  the cross-variations between the output and the
  contractible risk sources.
Thus,   like sample Sharpe ratios used in practice, path-dependent contracts naturally arise when there is  moral hazard with respect to
risk management. In a numerical example, we show that the loss of efficiency can be significant if the principal does not use
the quadratic variation component of the optimal contract.


\medskip






\section{Introduction}
\setcounter{equation}{0} \setcounter{Assumption}{0}
\setcounter{Theorem}{0} \setcounter{Proposition}{0}
\setcounter{Corollary}{0} \setcounter{Lemma}{0}
\setcounter{Remark}{0}
\setcounter{Definition}{0}\setcounter{Example}{0}

In many cases managers are in charge of managing exposures to many different types of risk,
and they do that dynamically. A well-known example is the management of a portfolio
of many risky assets.
Nevertheless, virtually all existing continuous-time principal-agent models
with moral hazard and continuous output value process suppose that the agent controls the drift
of the output process, and not its volatility components.
The drift is what the agent controls in  the seminal models of  Holmstrom and Milgrom (1987), henceforth
 HM (1987), in which  utility is drawn from
terminal payoff, and of Sannikov (2008), in which utility is drawn from inter-temporal payments.
In fact, in those papers the moral hazard cannot arise from volatility choice anyway when there is only one source of risk (one Brownian motion),
 because if the principal observes the output process
continuously, there is no  moral hazard with respect to
volatility choice: the volatility can be deduced from the output's quadratic variation process.
However, when the agent manages many non-contractible sources of risk, his choices of exposures to the individual risk sources
cannot be deduced from the output observations, even continuous.

\vspace{0.5em}
\noindent One reason this problem has not been studied is
the previous lack of a workable mathematical
methodology to tackle it. When the drift of an It\^o process is picked by the agent, this can be formulated as
a Girsanov change of the underlying probability measure to an equivalent probability measure, and there is an extensive mathematical theory
behind it.
However, changing volatility components requires singular changes of measures, a problem that, until recently, has not been
successfully studied. We take advantage of  recent progress in this regard, and use the new theory to
analyze our principal-agent problem. However, we depart from the usual modeling assumption
that the agent's effort consists in changing the distribution of the output (i.e., the underlying probability  measure),
and we, instead, apply the standard stochastic control approach in which the agent actually changes the values of the
controlled process, while the probability measure stays the same.  The reason why all the existing literature uses the
former, so-called ``weak formulation", is that the agent's problem  becomes tractable {\it for any given contract}.
Instead, we make  the agent's problem tractable  by restricting the family of admissible contracts to a natural set of contracts
that lead to a tractable characterization of the agent's problem. Essentially, we restrict the  admissible contracts to those
for which the agent's problem is solvable. It could be argued, that, from a practical point of view, these are the only relevant contracts --
the others, for which the principal does not know what incentives they will provide to the agent, will likely not be offered.
Moreover, we show that in the classical Holmstrom-Milgrom model
the restricted family is not actually restricted at all. Thus, in addition to solving the new agency problem with volatility control,
we offer an alternative new way to study classical problems with drift control only.

\vspace{0.5em}
\noindent What we have just described  is  our main contribution on the methodological side.
In principle, our approach can be applied to any utility functions. However, with terminal payment only, as in HM (1987), the only
tractable case is the one with CARA (exponential) utility functions. Our main economic insight of the paper is as follows.
The optimal contract is linear (in the CARA case), but not only in the output process as in HM (1987), rather, also in these factors:
the output, its quadratic variation, the contractible sources of risk (if any),
and the cross-variations between the output and the contractible risk sources.
Thus, the use of path dependent  contracts naturally arises when there is  moral hazard with respect to
risk management. In particular, our model is consistent with the use of the sample Sharpe ratio when compensating
 portfolio managers. However, unlike the typical use of Sharpe ratios, there are parameter values for which
 the principal rewards the agent for higher values of quadratic variation, thus, for taking higher risk.

\vspace{0.5em}
\noindent In case there are two sources of risk, and at least one is observable and contractible,
the first best is
attained, because there are two risk factors and at least two contractible variables, the output and at least one risk source;
however, to attain the first best, the optimal contract  makes use of
the quadratic and cross-variation factors.
In case of two non-contractible sources of risk, we solve numerically  a CARA example with a quadratic
cost function.
 In this case, first best is generically not attainable.
  Numerical computations show that the loss in expected utility can be significant if
  the principal does not use the path-dependent components of the optimal contract.

\vspace{0.5em}
\noindent
{\bf Literature review.}
An early continuous-time paper on volatility moral hazard
is Sung (1995). However, in that paper moral hazard is a result of  the output being
observable only at the terminal time, and not because of multiple sources of risk.
Consequently, the optimal contract is still a  linear function of the terminal output value only.
 The paper Ou-Yang (2003) shows that
the optimal contract  depends on the final value of the output
and a ``benchmark" portfolio, in an economy in which all the sources of risk (all the risky assets
available for investment) are observable, but the output is observable at final time only.
 Some of his results are extended in Cadenillas, Cvitani\'c and
Zapatero (2007), who show that, if the market is complete,  first-best is attainable  by contracts
that depend only on the final value of the output. Thus, second best may be different from first
best only if the market is not complete. In our model, the principal observes the whole path of the output, but not
all the exogenous sources of risk, and also there is a non-zero cost of effort, which makes the market incomplete. Thus,
    first best and
second best are indeed different for generic parameter configurations.

\vspace{0.5em}
\noindent
More recently, Wong (2013) considers the moral hazard of risk-taking in a model different from ours:
the horizon is infinite, as in Sannikov (2008), and, while the volatility is fixed,  the agent's effort influences
the arrival rate of Poisson shocks to the output process.
Lioui and Poncet (2013), like us, consider a principal-agent problem in which
the volatility is chosen by the agent.
Theirs is  the  first-best framework;  however, unlike  the above mentioned  papers,
 they assume that the agent has enough bargaining power to require that the contract be linear in the output and in  a benchmark factor.
  Working paper Leung (2014) proposes a model in which volatility moral hazard arises because
 there is an exogenous factor multiplying the (one-dimensional) volatility choice of the agent, and that  factor is not observed by the principal.
 In terms of  methodological techniques, a number of papers in mathematics literature has been developing tools for comparing
 stochastic  differential systems corresponding to differing volatility structures. We cite them in the main body of the paper, as we
 use a lot of their results.
 Here, we only mention  two papers that not only contribute to the development of those tools, but also apply them to
 problems in financial economics: Epstein and Ji (2013) and (2014). Theirs is not the principal-agent problem, but the
 ambiguity problem, that is, a model in which the decision-maker has multiple priors on the drift and the volatility  of
 market factors. There is  no  ambiguity in our model, it is  the agent who controls  the drift and the volatility of the output
 process.

\vspace{0.5em}
\noindent
We  start by Section 2  presenting the simplest possible example in our context, with CARA utility functions  and quadratic penalty,
 we describe  the general  model in Section 3, in Section 4 we present the contracting problem and our approach to solving it,
we consider the case with no exogenous contractible factors  in Section 5, and conclude with Section 6. The longer proofs are provided
in Appendix.

\section{Example: Portfolio Management with CARA Utilities and quadratic cost}
\setcounter{equation}{0} \setcounter{Assumption}{0}
\setcounter{Theorem}{0} \setcounter{Proposition}{0}
\setcounter{Corollary}{0} \setcounter{Lemma}{0}
\setcounter{Remark}{0}
\setcounter{Definition}{0}\setcounter{Example}{0}

As an illustrative tractable example,
we present here the Merton's portfolio selection problem
with two risky assets, $S_{1}$ and $S_{2}$, and a risk-free asset
with the continuously compounded rate set equal to zero.
Holding amount $v_{i}$ in asset $i$, the portfolio value process $X_t$ follows the dynamics
$$
dX=\frac{v_{1}}{S_{1}}d S_{1}+\frac{v_{2}}{S_{2}}dS_{2}.$$

\noindent Suppose
$$
dS_{i,t}/S_{i,t}=b_{i}dt+dB^i_{t},$$
where $B^i$ are independent Brownian motions and $b_i$ are constants.
We have then
$$
dX_t=\left [v_{1,t}b_1+v_{2,t}b_2\right]dt + v_{1,t}dB^1_t+v_{2,t}dB^{2}_t.$$
The principal hires an agent to manage the portfolio, that is, to choose the values of $v_t=(v_{1,t}, v_{2,t})$, continuously in time.
 We assume that the agent is paid only at the final time $T$ in the
amount $\xi_T$.  The utility of the principal is
$
 U_P(X_T-\xi_T)$
and
the utility of the agent is
 $
  U_A(\xi_T- K^{v}_{T})$ where
  $K_T^{v}:=\int_0^Tk(v_s)ds$ and
   $k:\mathbb R^2\rightarrow \mathbb R$ is a non-negative convex cost function.

\vspace{0.5em}
\noindent
If the principal only observes the path of $X$, then she can deduce the value of $v_{1,t}^2+v_{2,t}^2$,
but not the values of $v_{1,t}$ and $v_{2,t}$ separately, which leads to  moral hazard. On the other hand, if she also observes the price path of one of the assets, say
$S_1$, then she can deduce the (absolute) values of  $v_{1,t}$ and of $v_{2,t}$.
To distinguish between these cases, denote by ${\bf 1}_O$ the indicator function that is equal to one if the path of $S_1$ is observable and contractible, and zero otherwise.

\vspace{0.5em}
\noindent We assume that the principal and the agent have exponential utility functions,
 $$U_P(x)=-e^{-R_Px}~,~~U_A(x)=-e^{-R_Ax}.$$
 and that the agent's running cost of portfolio $(v_1,v_2)$ is of the form $$k(v_1,v_2)=\frac{1}{2}\beta_{1}(v_{1}-\alpha_{1})^{2}+\frac{1}{2}\beta_{2}(v_{2}-\alpha_{2})^{2}.$$
 Thus, it is costly to move the volatility $v_i$ away from $\alpha_i$,
and the cost intensity is $\beta_i$. An interpretation is that $\alpha_i$ are the
initial risk exposures of the firm at the time the manager starts his contract.

\subsection{First-best contract}

Given a ``bargaining-power" parameter $\rho>0$,  the principal's first-best problem is defined as
 $$
 \underset{v}{\sup} ~ \underset{\xi_T}{\sup}\ \mathbb E\left[U_P(X_T-\xi_T)+\rho U_A(\xi_T-K_T^v)\right].$$
 The first order condition for $\xi_T$ is then
$$U_P'(X_T-\xi_T)=\rho U_A'(\xi_T-K^{v}_T).$$
With CARA utilities, we obtain
$$\xi_T=\frac{1}{R_A+R_P}\left(R_P X_T+R_AK^{v}_T+ \log\left(\frac{\rho R_A}{R_P}\right)\right).$$
Thus, the optimal first best contract is linear in the final value $X_T$ of the output. Plugging back into the optimization problem, we get that it is equivalent to
\begin{align*}
& -C_\rho\underset{{v}}{\inf} ~\mathbb E\left[\exp\left(-\frac{R_AR_P}{R_A+R_P}\left(X_T-K^{v}_T\right)\right)\right]\\[0.5em]
&=-C_\rho\underset{v}{\inf}~ \mathbb E\left[\mathcal E\left(-\frac{R_AR_P}{R_A+R_P}\int_0^Tv_s\cdot dX_s\right)\exp\left(-\frac{R_AR_P}{R_A+R_P}\int_0^Tf(v_s)ds\right)\right],
\end{align*}
where $x\cdot y$ denotes the inner product, $C_\rho$ is a constant,
 $\mathcal E$ denotes the  Dol\'eans-Dade stochastic exponential\footnote{Stochastic exponential is defined by
 $$
  \mathcal  E\left(\int_t^uX_sdB_s\right)
  =
  e^{-\frac12\int_t^u\abs{X_s}^2ds+\int_t^uX_sdB_s}.
  $$ }, and
 \begin{eqnarray*}
 f(v):=b\cdot v-k(v)-\frac12\frac{R_AR_P}{R_A+R_P}\No{v}^2.
 \end{eqnarray*}
Under technical conditions,
Girsanov theorem can be applied
and the above can be written as
$$ -C_\rho\underset{v}{\inf}~ \mathbb E^{\widehat{ \mathbb P}}\left[
\exp\left(-\frac{R_AR_P}{R_A+R_P}\int_0^Tf(v_s)ds\right)\right].$$
for an appropriate probability measure $\widehat{ \mathbb P}.$
Thus, first best optimal $v=v^{FB}$ is deterministic, found by the pointwise maximization of the function $f(\cdot)$,
  annd given by\footnote{In the case the cost $\beta_i$ is zero, the first best volatility  $v_i^{FB}$ is simply a product of  the risk premium $b_i$
and the aggregate prudence $1/\bar{R}$.}
\begin{eqnarray}\label{firstbest}
v_i^{FB}:=\frac{b_i+\beta_i\alpha_i}
                        {\beta_i+\bar R},
 &\mbox{where}&
 \frac{1}{\bar R}
 :=
 \frac{1}{R_A}+\frac{1}{R_P}.
 \end{eqnarray}

\subsection{Second best contracts}

We now take into account that it is not the principal who controls $v$, but the agent.
We consider linear contracts based on the path of the observable portfolio value $X$, the observable quadratic variation of $X$, and,
possibly, on $S_1$ via $B^1$, and the
co-variation of $X$ and $B_1$. More precisely, let
\begin{equation}\label{lin2}
\xi_{T}=\xi_{0}+\int_{0}^{T}\left[Z^X_{s}dX_{s}+Y^X_{s}d\langle X\rangle_{s}+{\bf 1}_O\left(Z^1_sdB^1_s+Y^{1}_{s}d\langle X,B_1\rangle_{s}\right)+H_{s}ds\right],
\end{equation}
for some constant $\xi_0$, and some adapted processes $Z^X,Z^1,Y^X,Y^1$ and $H$.
To be consistent with the notation in the general theory that follows later, for future notational convenience we
work instead with arbitrary adapted processes $Z^X,Z^1,\Gamma^X,\Gamma^1$ and $G$ such that
\begin{align*}
Y^X&=\frac12 \left(\Gamma^X+ R_A(Z^X)^2\right),\\
Y^1&=\Gamma^1+R_AZ^XZ^1,\\
H&=-G+\frac 12 R_A(Z^1)^2.
\end{align*}


\noindent Solving the agent's problem if those processes were deterministic would be easy, but not for arbitrary choices of those processes.
We will allow them to be stochastic, but we will restrict the choice of  process $G_t$, motivated by a stochastic control analysis of the agent's problem,
discussed in a later section.  We will see in that section that the natural choice for $G_t$ is  $G_t:=G(Z^X_t,Z^1_t,\Gamma^X_t,\Gamma^1_t)$, where
\begin{align}\label{G_t}
\nonumber  G(Z^X,Z^1,\Gamma^X,\Gamma^1) &:= \sup_{v_{1},v_{2}}g(v_{1},v_{2},Z^X,Z^1,\Gamma^X,\Gamma^1)\\ &:= \sup_{v_{1},v_{2}} \left\{-k(v_{1},v_{2})+\frac12\Gamma^X(v_{1}^2+v_{2}^2)+Z^Xb\cdot v+{\bf 1}_O \Gamma^1 v_{1}
\right\}.
\end{align}
One of our main results, Theorem \ref{th:reg}, characterizes  the contracts that have a representation
of the above form, with $G_t$ as defined here. In particular, the theorem will show that applying the same method to the classical Holmstrom-Milgrom (1987)
problem leads to no loss of generality in the choice of possible contracts (including nonlinear contracts), and we will argue that, in general, this choice of contracts includes all practically relevant contracts.

\vspace{0.5em}
\noindent The agent is maximizing $-\mathbb E[-\exp(-R_A(\xi_T-K_T^v)]$ with $\xi_T$ as in \reff{lin2}.
%
This turns out to be an easy optimization problem:
 by Girsanov theorem (assuming appropriate technical conditions), similarly as in the first best problem,
 the agent's objective  can be written as
$$-\mathbb E_t^{{ \mathbb P}^*}\left[e^{-R_{A}\int_{0}^{T}\left[g_s-G_s\right]ds}\right],
$$
for an appropriate probability ${\mathbb P}^*$.
With our definition of $G$, we see that this is never larger than minus one, and it is equal to minus one for any pair $(v_1^*(s),v_2^*(s))$ (if it exists) that maximizes $g_s:=g(Z^X_s,Z^1_s,\Gamma^X_s,\Gamma^1_s)$,
$s$ by $s$, and $\omega$ by $\omega$. Thus, the agent would choose one of such pairs.

\vspace{0.5em}
\noindent
Notice that this implies that the principal can always make the agent  indifferent about one of the portfolio positions.
For example, if $b_2$ is not zero, she can set
$Z^X=-\alpha_2\beta_2/b_2$, and
$\Gamma^X\equiv \beta_2$, to make $g$ independent of $v_2$, hence the agent indifferent with respect to $v_2$. This is also possible if there is only one stock, say $S_2$,
so that in that case the first best is attained with such a contract,
 if we assume that the agent will choose what is best for the principal, when indifferent.

\subsubsection{Contractible $S_1$: first best is  attained}
With two  risky assets, if $S_1$ is observed, then also the covariation between $S^1$ and $X$ is observed, which means that $v_1$ is observed.
Since also the quadratic variation is observed, then $|v_2|$ is observed, and, if the observed processes are contractible,  we would expect the first best to be attainable.
Indeed,  $(v^*_1,v_2^*)$ is obtained by maximizing
$$g=-\frac12\beta_1(v_1-\alpha_1)^2-\frac12\beta_2(v_2-\alpha_2)^2+Z^X  b\cdot v+\Gamma^{1} v_1+\frac12\Gamma^X \|v\|^2
+Z^1b_1+(Z^1)^2+2Z^1Z^Xv_{1}.$$
Assume, for example, that $b_2\neq 0$,  $\beta_2\le\beta_1$.
Suppose the principal sets
\begin{align*}
\Gamma^X_t&\equiv \beta_2,\\
Z^X_t&\equiv -\alpha_2\beta_2/b_2,\\
\Gamma^{1}_t&=-\alpha_1\beta_1-Z^X_tb_1+
(\beta_1-\beta_2)v_1^{FB},\\
Z_1&\equiv 0.
\end{align*}
Then,
$$g=(\beta_2-\beta_1)\left[\frac12v_{1}^2-v_{1}v_1^{FB}
\right]+const.$$
We see that the agent is indifferent with respect to which $v_2$ he applies, and he would choose
$v_1^*=v_1^{FB}$.
Thus,
 if, when indifferent, the agent will choose
what is best for the principal, he will choose the first best actions. It can also be verified that the principal will attain the first
best expected utility with this contract.

\subsubsection{Non-contractible $S_1$}

If $S_1$ is not contractible,
the optimal   $(v^*_1,v_2^*)$ is obtained by maximizing
$$g(v_1,v_2)=-\frac12\beta_1(v_1-\alpha_1)^2-\frac12\beta_2(v_2-\alpha_2)^2+Z^X  b\cdot v+\frac12\Gamma^X \|v\|^2.$$
Assume, for example,  $\beta_2\le\beta_1$.
If $\Gamma^X>\beta_2$, then the agent will optimally choose $|v_2^*|=\infty$.
 It is straightforward to verify
 that this cannot be optimal for the principal.
The same is true  if  $\Gamma^X=\beta_2$ and $Z^X$ is not equal to $(-\alpha_2\beta_2)$.
If  $\Gamma^X=\beta_2$ and $Z^X=-\alpha_2\beta_2$, then the agent is indifferent with respect to which $v_2$ to choose.
If $\Gamma^X<\beta_2\le \beta_1$, the optimal positions are
$$v_i^*= \frac{Z^Xb_{i}+\alpha_{i}\beta_{i}}{\beta_i-\Gamma^X}.$$

\noindent
The principal's utility
is proportional to
\[
-\mathbb E\left[e^{-R_{P}\left(\int_{0}^{T}\left[(1-Z^X_{s})dX_{s}-\frac{1}{2}\Gamma^X_{s}d\langle X\rangle_{s}+G_{s}ds\right]\right)}\right].
\]
With the above choice of $v_i^*$, by a similar Girsanov change of probability measure as above, it is straightforward to verify that maximizing this is the same as maximizing, over $Z=Z^X$ and $\Gamma=\Gamma^X$,
\[
b\cdot v^*(Z,\Gamma)-\frac{1}{2}[R_AZ^2+R_P(1-Z)^2]\|v^*\|^2-k(v^*(Z,\Gamma)).
\]
This is a problem that can be solved numerically.\footnote{In Appendix, we provide sufficient conditions for  existence of optimizers.
Numerically, we found optimizers for all  parametric choices we tried.}
We now present a numerical example that will show us first, that first best is not attained,
  second, that the optimal contract contains a non-zero quadratic variation component and that ignoring it
can lead to substantial  loss in expected utility, and third, that there are parameter values for which
the principal rewards the agent for taking high risk (unlike the typical use of portfolio Sharpe ratios in practice).

\vspace{0.5em}
\noindent
In Figure 1 we plot the percentage loss in the principal's second best utility certainty equivalent
relative to the first best, when varying the parameter $\alpha_2$, and keeping everything else fixed.
The loss can be significant for extreme values of  initial exposure $\alpha_2$.
That is, when the initial risk exposure is far from desirable, the moral hazard cost of
providing incentives to the agent to modify the exposure is high.
  \begin{center}
\includegraphics[scale=0.23]{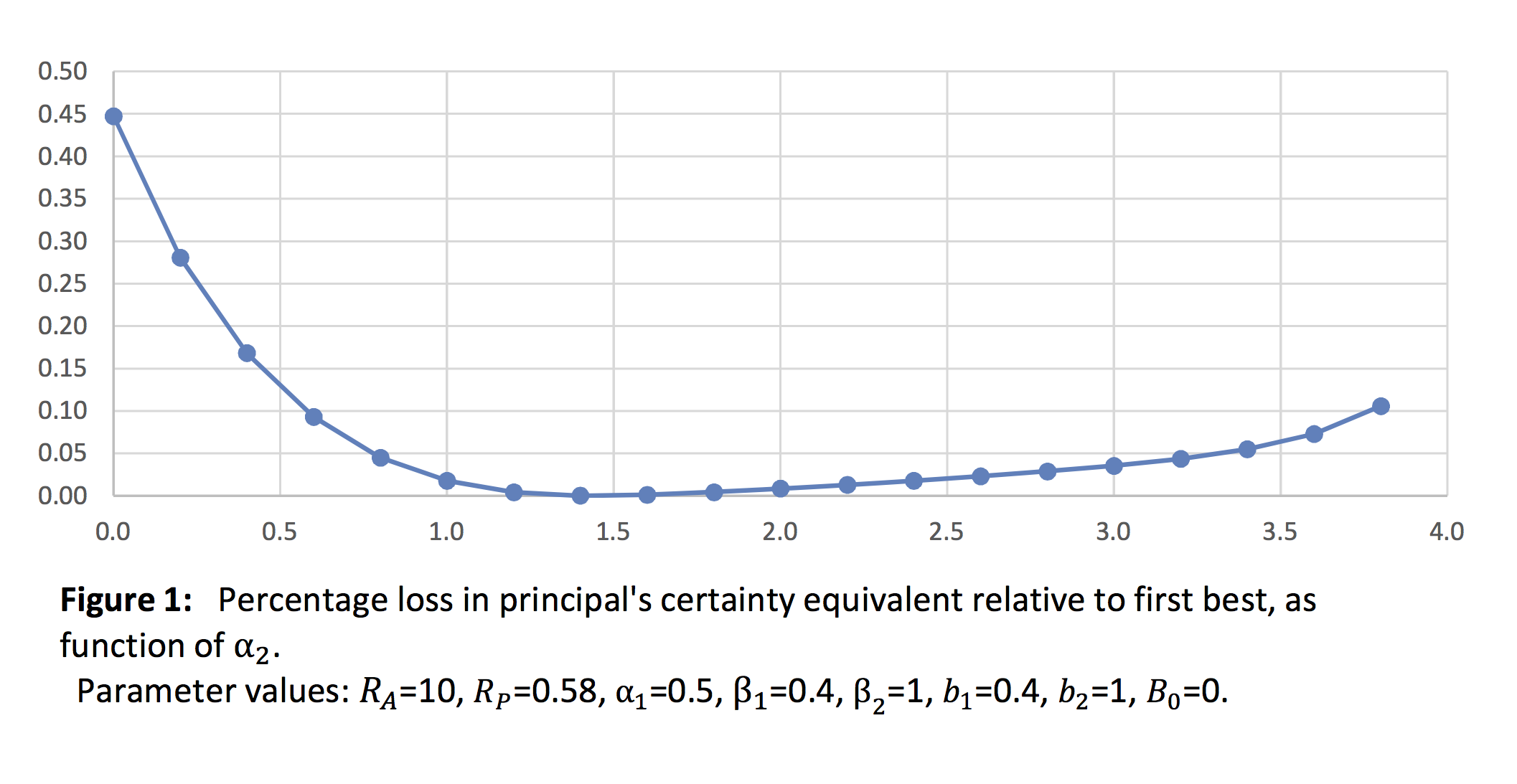}
\end{center}

\vspace{-2em}

\noindent
In Figure 2, we compare the principal's second best certainty equivalent to the one she would obtain if offering the
contract that is optimal among those that are linear in the output, but do not depend on its quadratic variation.
Again we see that  the corresponding relative percentage loss can be large.

  \begin{center}
\includegraphics[scale=0.23]{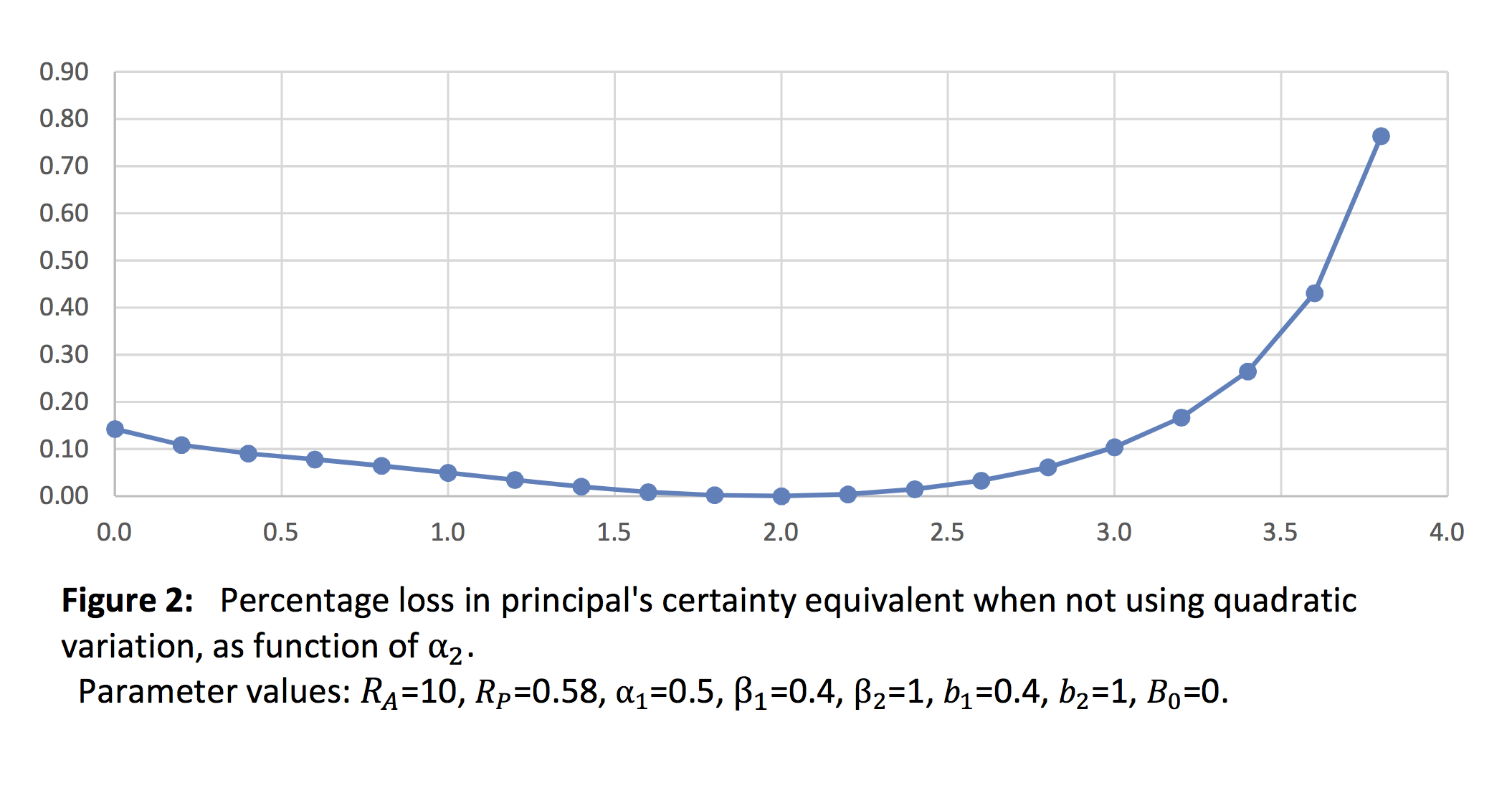}
\end{center}
\vspace{-2em}

\noindent
Figure 3 plots the values of the coefficient (the sensitivity) multiplying the quadratic variation in the optimal contract. We see that the principal uses  quadratic variation as an incentive tool: for low values of the initial risk exposure $\alpha_2$
she wants to increase the risk exposure by rewarding
higher variation (the sensitivity is positive), and for its high values she wants to decrease it
by penalizing high variation (the sensitivity is negative).
 This is because when the initial risk exposure $\alpha_2$ is not at  the desired value $v_2^*$,
 incentives are needed to make the agent apply costly effort to modify the exposure.
\vspace{-2em}

  \begin{center}
\includegraphics[scale=0.23]{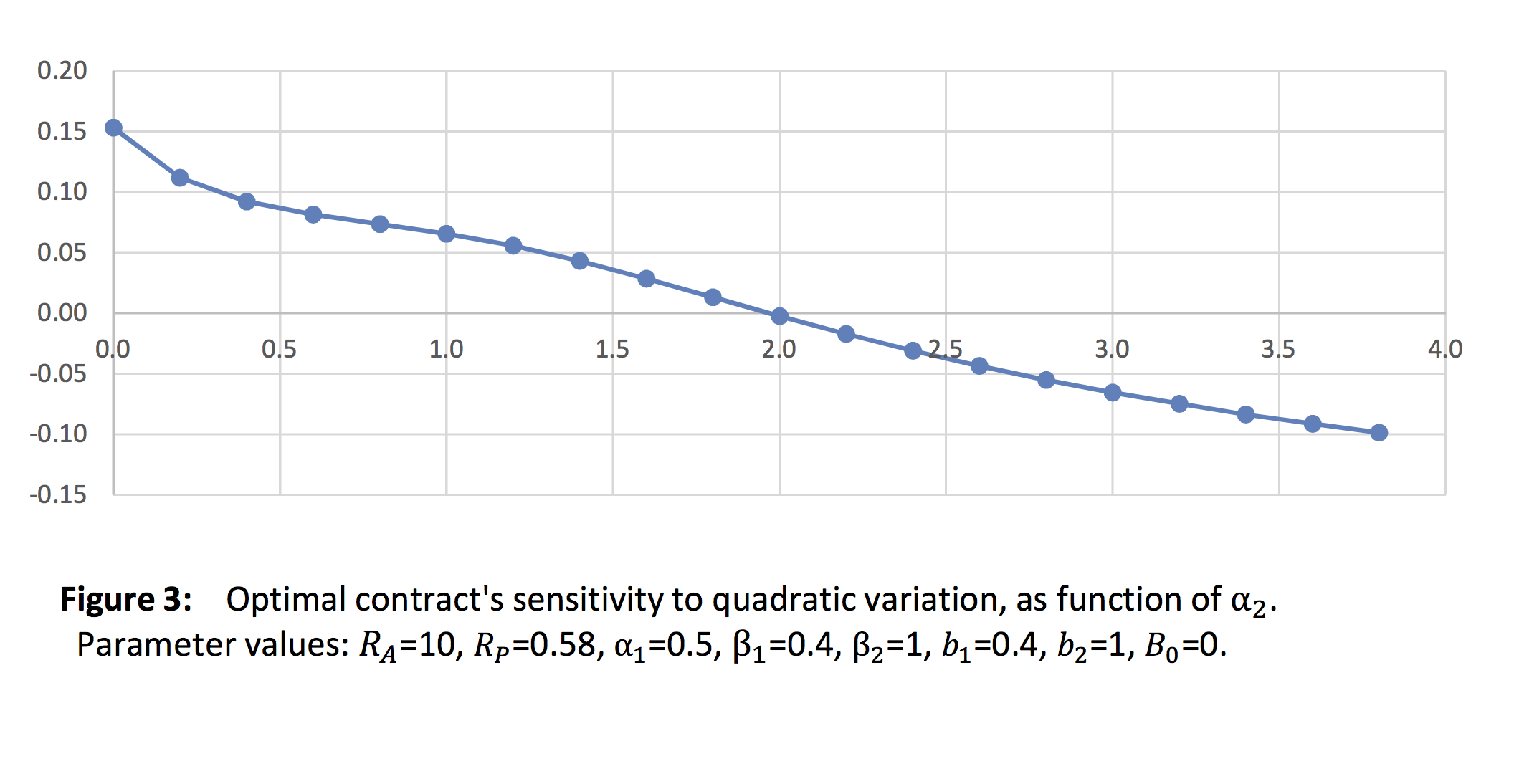}
\end{center}
\vspace{-2em}
\noindent
In the rest of the paper, we generalize this example  and
we aim to characterize the contract  payoffs that can be represented as in (\ref{lin2}), with $G_t$ defined analogously to
(\ref{G_t}).

\section{The General Model}
\setcounter{equation}{0} \setcounter{Assumption}{0}
\setcounter{Theorem}{0} \setcounter{Proposition}{0}
\setcounter{Corollary}{0} \setcounter{Lemma}{0}
\setcounter{Remark}{0}
\setcounter{Definition}{0}\setcounter{Example}{0}


We consider the following general  model for the output process $X^{v,a}$:
\begin{equation}\label{weak0}
X^{v,a}_t=\int_0^t \sigma_s(v_s)\cdot(b_s(a_s)ds+ dB_s),
\end{equation}
where $(v,a)$ represents the control pair of the agent, allowing also for separate control $a$ of  the drift, where $v$ and $a$ are adapted processes taking values in some subset $\mathcal V\times\mathcal A$ of $\mathbb R^m\times\mathbb R^n$, for some $(m,n)\in\mathbb N^*\times\mathbb N^*$. Moreover, $b:[0,T]\times\mathcal A\longrightarrow \mathbb R^d$ and $\sigma:\mathcal V\longrightarrow \mathbb R^d$ are given deterministic functions such that
\begin{equation}\label{eq:lingrowth}
\No{b(a)}+\No{\sigma(v)}\leq C\left(1+\No{v}+\No{a}\right),
\end{equation}
for some constant $C>0$, and $B=(B^1,\ldots,B^d)$ is a $d$-dimensional Brownian motion, and the products are inner products of vectors, or a matrix acting on a vector.

\vspace{0.5em}
\noindent The example of delegated portfolio management corresponds to the case in which  $m=n=d$, and
\begin{equation}\label{eq:simple}
\sigma_t(v):=v^T\sigma,\ b_t(a):=b,
\end{equation}
for some fixed $b\in\mathbb R^d$ and some invertible $d\times d$ matrix $\sigma$.
 In that case, the interpretation of the process $X^{v,a}$
is that of the portfolio value  process dependent on the agent's choice of the  vector $v$ of portfolio dollar-holdings in $d$ risky assets with  volatility matrix $\sigma$,  and the vector $b$ of risk premia.  Another special case, when $d=1$, $v$ is fixed and the agent  controls $a$ only,
 is the original continuous-time principal-agent model of Holmstrom and Milgrom (1987).%

\vspace{0.5em}
\noindent
In addition to the output process $X^{v,a}$, we may want to allow contracts based on additional observable and contractible  risk factors $B^1,\ldots,B^{d_0}$,
 for some  $1\le d_0 < d$. For example, $S_t=S_0+\mu_S t +  \sigma_SB^1_t$ might be a model for a contractible stock index.


\vspace{0.5em}
\noindent
Usually in contract theory, for sake of tractability, the model  is considered in its so-called weak formulation, in which  the agent changes the output process not by changing directly the controls $(v,a)$, but by changing the probability measure over the underlying probability space. When, as in standard continuous-time contract models, the effort is present only in the drift, changing measures is done by the means of the Girsanov theorem.
Until recently, though, such a tool had not been available for singular changes of measure
that are needed when changing volatility, as is the case here. The mathematics to formulate rigorously the weak formulation of our problem is now available.\footnote{We used the weak formulation in an earlier version of this paper.} However,  we take a different approach: instead of assuming the weak formulation, we will
adopt the standard strong formulation of  stochastic control (no changes of measure). We will still be able to explicitly characterize the solution to  the agent's problem, but only in a restricted family of admissible contracts.
We will provide an assumption under which a contract belongs to the restricted family, we will argue that the family includes all
 contracts relevant in practice, and we will show the assumption is not restrictive  if $v$ is fixed. Thus, we also provide a new alternative way to solve the Holmstrom-Milgrom (1987) problem.
 \footnote{Interestingly,  to derive these results, even though we work with the strong formulation, we  need to use the weak formulation in the proofs. In fact, for the admissible contracts in our restricted family, the weak and the strong formulation for the agent's problem are  equivalent.}

\vspace{0.5em}
\noindent We first need to
introduce some notation and the framework. We work on the canonical space $\Omega$ of continuous functions on $[0,T]$, with its Borel $\sigma$-algebra $\mathcal F$.
The
$d-$dimensional
canonical process is denoted $B$, and $\mathbb F:=(\mathcal F_t)_{0\leq t\leq T}$ is its natural filtration.
Let $\mathbb P_0$ denote  the
 $d-$dimensional
  Wiener measure on $\Omega$. Thus, $B$ is a $d-$dimensional Brownian motion under $\mathbb P_0$.
    We denote by $\mathbb E$ the expectation operator under $\mathbb P_0$.

\vspace{0.5em}
\noindent
A pair $(v,a)$ of $\mathbb F$-predictable processes taking values in $\mathcal V\times\mathcal A$ is said to be admissible if
\begin{equation}\label{eq:expmom}
\int_0^T|\sigma_s(v_s)\cdot b_s(a_s)|<+\infty,\ \mathbb P_0-a.s.,\ \mathbb E\left[\exp\left(p\int_0^T||\sigma_s(v_s)||^2ds\right)\right]<+\infty,\ \text{for all $p>0$,}
\end{equation}
and
\begin{equation}\label{eq:martexp}\mathcal E\left(\int_0^\cdot b_s(a_s)\cdot dB_s\right)\text{ is a $\mathbb P_0$-martingale in $L^{1+\eta}$, for some $\eta>0$,}
\end{equation}
where
  \begin{equation}\label{eq:invertible}
  \No{\sigma_s(v_s)}^2-\sum_{i=1}^{d_0}\abs{\sigma_s^i(v_s)}^2\neq 0.
  \end{equation}
  Moreover, we assume that the first $d_0$ entries  of vector $b$ do not depend on the control process $a$ (because they will correspond to
 exogenous  contractible factors). 
 \footnote{The above integrability conditions are suitable for the case of CARA utilities,and they may have to be modified for other utility functions.
 We discuss briefly the general case later below.
 Notice also that condition \eqref{eq:invertible} in the case of portfolio management problem  with $a=0$ and $\sigma=I_d$,  means that the investor has to invest in at least one of the non-contractible sources of risk. When there are no contractible source of risk (that is when $d_0=0$), this condition reduces to $\sigma_s(v_s)\neq 0_{1,d}$.}
 \begin{Remark}
The condition  \reff{eq:invertible} is actually not needed to prove our results; it is  only needed 
in our  computations that motivate the definition of the admissible contracts. 
 \end{Remark}

\vspace{0.5em}
\noindent As in the example in the previous section, we assume that the agent is paid only at the final time $T$ in the
amount $\xi_T$. When the agent chooses the controls $(v,a)\in\mathcal U$, the utility of the principal is
$
 U_P(X_T^{v,a}-\xi_T)$, and the
 utility of the agent is  $-e^{-R_A (\xi_T-K^{v,a}_{0,T})}$ where
  $$K^{v,a}_{t,T}=\int_t^Tk(v_s,a_s)ds.$$
  and $k:\mathbb R^m\times\mathbb R^n\longrightarrow \mathbb R$ is a non-negative (convex) cost function.



\section{Second Best with Contractible  Risks}

In this section, we assume there is exactly one exogenous contractible risk factor, that is, we  set $d_0=1$\footnote{We could equally have any subset of $\{B^{1},\ldots, B^{d}\}$ contractible, and the rest non-contractible.
We assume that there is only one contractible risk source for simplicity of notation, and because it is also consistent with the ``systemic risk -- stock index" interpretation. We  consider the case in which none of the risk sources is contractible in the following section.}.
Thus,  we interpret $B^1$ as the observable and contractible systemic risk.

\subsection{Setup}\label{sec:sec}


 We allow the contract payoff to depend both on the output $X^{v,a}$ and $B^1$. That is, given a pair $u:=(v,a)$ chosen by the agent, the principal can offer contract payoffs measurable with respect to  ${\cal F}^\obs_T$, a $\sigma$-field contained in the filtration
  $\mathbb F^\obs:=\mathbb F^{X^{v,a}}\vee\mathbb F^{B^{1}}$ generated by $(X^{v,a},B^1)$,
  where $\mathbb F^{X^{v,a}}:=\{{\cal F}^{X^u}_t\}_{0\leq t\leq T}$ is the (completed) filtration generated by the
  output process $X^u$. Recall that our assumptions imply that  $b^1$ does not depend on $a$, and  introduce the following notation for the contractible and non-contractible factors:
    $$B^{\obs,v,a}:=(
  X^{v,a},B^1)^T=\int_0^\cdot\mu_s(v_s,a_s)ds+\int_0^\cdot\Sigma_s(v_s)dB_s,
  $$ $$ B^{\nobs,v,a}:=\int_0^\cdot\Sigma^\perp_s(v_s)dB_s,$$
where for any $(v,a)\in\mathcal V\times\mathcal A$ and any $s\in[0,T]$, the $\mathbb R^2$ vector $\mu_s(v,a)$ and the $2\times d$ matrix $\Sigma_s(v)$ are defined by
$$ \mu_s(v,a):=\begin{pmatrix}\sigma_s^T(v)b_s(a)\\ 0\end{pmatrix},\ \Sigma_s(v):=\begin{pmatrix}\displaystyle \sigma_s^T(v)\\
I_{1,d}
\end{pmatrix},\text{Êwith } I_{1,d}:=
\begin{pmatrix}1 & 0_{1,d-1}\end{pmatrix},$$
and where  $0_{p,q}$ denotes the  $p\times q$ matrix of zeros. Furthermore, $\Sigma^\perp_s$ is a $(d-2)\times d$ matrix satisfying, for any $v\in\mathcal V$ and any $s\in[0,T]$,
$$\Sigma_s^\perp(v)\Sigma_s^T(v)=0_{d-2,2}\ \text{ and } \ \Sigma^\perp_s(v)\left(\Sigma^\perp_s\right)^T(v)=I_{d-2}.$$
In other words, we take as the non-contractible sources of risk  everything which is "orthogonal" to the contractible sources of risk.

\vspace{0.5em}
\noindent Note that the vector $\begin{pmatrix}B^{\obs,v,a} & B^{\nobs,v,a}\end{pmatrix}^T$ generates the same filtration $\mathbb F$ as $B$ if and only if  the density of its quadratic variation is invertible. Using the definition of $\Sigma^\perp$ and $\Sigma$, we can readily compute that
$$\frac{d\Big\langle\begin{pmatrix}B^{\obs,v,a}&B^{\nobs,u}\end{pmatrix}^T\Big\rangle_s}{ds}=\begin{pmatrix}\Sigma_s(v_s)\Sigma_s^T(v_s) & 0_{2,d-2}\\
0_{d-2,2}& I_{d-2}\end{pmatrix},$$
which is invertible if and only if $\Sigma_s(v_s)\Sigma_s^T(v_s)$ is itself invertible. Simple calculations lead us  to the following necessary and sufficient condition
\begin{equation*}
\No{\sigma_s(v_s)}^2-\left(\sigma_s^1(v_s)\right)^2\neq 0,
\end{equation*}
which is exactly \eqref{eq:invertible} when $d_0=1$, and thus explains why we assumed \eqref{eq:invertible}.

\subsection{Admissible contracts}

In this subsection we will define the set of admissible contract payoffs.
To motivate our definition, consider the agent's value function at time $t$, for a given choice of control $(v,a)\in\mathcal U$
\begin{equation}\label{agentvalue}
V_t^{A,v,a}:=\underset{(v',a')\in\mathcal U((v,a),t)}{\rm essup}\ \mathbb E_t\left[U_A(\xi_T-K^{v',a'}_{t,T})\right],
\end{equation}
where the set $\mathcal U((v,a),t)$ denotes the subset of elements of $\mathcal U$ which coincide with $(v,a)$, $dt\times\mathbb P_0-$a.e. on $[0,t]$.
 Note that  we have the following explicit relationship between the payoff and the terminal value of the value function:
 \begin{equation}\label{xiV}
 \xi_T=
U_A^{-1}(V^{A,v,a}_T\big).
 \end{equation}
 The idea  is to consider the most general representation of
 the value function $V^A$ that we can reasonably expect to have, and then define the admissible
 contract payoffs via \reff{xiV}.
When $U_A$ is a CARA utility function,  we may guess from \reff{xiV} that contracts $\xi_T$
 are such that $\xi_T$ can be written as a linear combination of various integrals. For example, in the special case $d_0=0$ with no outside contractible factors,
 we might expect the contracts to satisfy
 $$U_A\left(\xi_T\right)=C\exp\left[-R_A\left(\int_0^TZ_sdX^{v,a}_s+\int_0^TY_sd\langle X^{v,a} \rangle_s +\int_0^T H_sds\right)\right],$$
 for some  constant $C< 0$ and some adapted processes $H, Y, Z$.
 With $U_A(x)=-e^{-R_Ax},$ this would give
 \begin{equation}\label{xisat}
 \xi_T= \tilde C+ \int_0^TZ_sdX^{v,a}_s+\int_0^TY_sd\langle X^{v,a} \rangle_s +\int_0^T H_sds,
  \end{equation} for some  constant $\tilde C$.
 We will admit exactly the contracts of this form, but, taking into account \eqref{xiV},  we will impose some restrictions on process
 $H$. We present this reasoning next, in an informal way, to motivate the definition which will follow thereafter.

\vspace{0.5em}
\noindent Notice that $V^{A,v,a}_t$ is an ${\cal F}_t-$measurable function and can thus be written in a functional form as
 \begin{eqnarray*}
 V^{A,v,a}_t
 &=&
 V^A\big(t,B^{\obs,v,a}_.,
                B^{\nobs,v,a}_.\big)
 \;=\;
 V^A\big(t,(B^{\obs,v,a}_s)_{s\le t},
                  (B^{\nobs,v,a}_s)_{s\le t}\big).
 \end{eqnarray*}

\noindent
If the value function is smooth in the sense of the Dupire (2009) functional differentiation (see also Cont and Fourni\'e (2013) for more details), then, the Dupire time derivative $\partial_tV^{A,v,a}$
exists, and one can find predictable processes
\begin{eqnarray*}
  \tilde Z^{v,a}=\left(\begin{array}{c}  \tilde Z^{\obs,v,a}
                                               \\
                                                \tilde Z^{\nobs,v,a}
                    \end{array}
             \right)
 &\mbox{and}&
 \tilde \Gamma^{v,a}=\left(\begin{array}{ccc} \tilde \Gamma^{\obs,v,a} & 0
                                                     \\
                                                     0 & \tilde \Gamma^{\nobs,v,a}
                        \end{array}\right),
 \end{eqnarray*}
with $ \tilde Z^{\obs,v,a},  \tilde Z^{\nobs,v,a}$ taking values in $\mathbb{R}^2$ and $\mathbb{R}^{d-2}$, respectively, $\tilde \Gamma^{\obs,v,a},\tilde \Gamma^{\nobs,v,a}$ taking values in the spaces $\mathbb{S}_2$ and  $\mathbb{S}_{d-2}$ of symmetric matrices, respectively, such that we have the following generalization of It\^o's rule (see Theorem 1 in Dupire (2009) or Theorem 4.1 in Cont and Fourni\'e (2013)),
 \begin{align}
\nonumber dV^{A,v,a}_t
 =&\
 \partial_tV^{A,v,a}_t
 +\frac12\mbox{Tr}\Big[\tilde \Gamma^{\obs,v,a}_t
                                    d\langle B^{\obs,v,a}\rangle_t\Big]
 +\frac12\mbox{Tr}\Big[\tilde \Gamma^{\nobs,v,a}_t
                                    d\langle B^{\nobs,v,a}\rangle_t\Big]
 +  \tilde Z^{\obs,v,a}_t
    \cdot dB^{\obs,v,a}_t
 \\
 \nonumber&+  \tilde Z^{\nobs,v,a}_t
    \cdot dB^{\nobs,v,a}_t
 \nonumber\\
 =&\
 \Big(\partial_tV^{A,v,a}_t
         +\frac12\mbox{Tr}\big[\tilde\Gamma^{\obs,v,a}_t
                                            \Sigma_t(v_t)\Sigma_t^T( v_t)
                                     \big]
         +\frac12\mbox{Tr}\big[\tilde\Gamma^{\nobs,v,a}_t\big]
         + \tilde Z^{\obs,v,a}_t\cdot\mu(v_t,a_t)
 \Big)dt
 \nonumber\\
 &
 +  \Big(\Sigma_t^T(v_t) \tilde Z^{\obs,v,a}_t
            + \left(\Sigma^\perp_t\right)^T(v_t) \tilde Z^{\nobs,v,a}_t
     \Big)\cdot dB_t, \ \mathbb P_0-a.s.
 \label{eq:eq}
 \end{align}
For instance, if we happen to be in the Markov case in which $V^{A,v,a}_t=f(t,B^{\obs,v,a}_t,B^{\nobs,v,a}_t)$ for some smooth function $f(t,x,y)$, 
then, it follows from It\^o's rule that the processes $ \tilde Z^{\obs,v,a}, \tilde Z^{\nobs,v,a}$ and $\tilde\Gamma^{\obs,v,a}, \tilde\Gamma^{\nobs,v,a}$ are given by
 \begin{eqnarray*}
 \tilde Z^{\obs,v,a}_t
 =\partial_xf(t,B^{\obs,v,a}_t,
B^{\nobs,v,a}_t),
 &&
 \tilde Z^{\nobs,v,a}_t
 =\partial_yf(t,B^{\obs,v,a}_t,
 B^{\nobs,u}_t),
 \\
 \tilde\Gamma^{\obs,v,a}_t
 =\partial_{xx}f(t,B^{\obs,v,a}_t,
                         B^{\nobs,v,a}_t),
 &&
\tilde \Gamma^{\nobs,v,a}_t
 =\partial_{yy}f(t,B^{\obs,v,a}_t,
B^{\nobs,v,a}_t),
 \end{eqnarray*}
with $\partial_x,\partial_{xx},\partial_y,\partial_{yy}$ denoting partial derivatives with respect to the corresponding variables.

\vspace{0.5em}
\noindent Next, from the martingale optimality principle of the classical stochastic control theory, the dynamic programming principle suggests that the process $V^{A,v,a}_te^{R_A K^{v,a}_{0,t}}$ should be a supermartingale for all  admissible controls $(v,a)$, and that it should be  a martingale for any optimal control $(v^*,a^*)$, provided such exists. Writing formally that the drift coefficient of a supermartingale is non-positive, and that of a martingale must vanish, we obtain the following path-dependent HJB (Hamilton-Jacobi-Bellman) equation:
 \begin{equation}\label{eq:HJB}
 -\partial_t V^{A,v,a}_t +R_AV_t^{A,v,a} G_1(t,Z^{v,a}_t,\Gamma^{v,a}_t)
 =
 0,\ \text{where}\ (Z_t^{v,a},\Gamma_t^{v,a}):=-\frac{1}{R_AV_t^{A,v,a}}\left(\tilde Z_t^{v,a},\tilde \Gamma_t^{v,a}\right),
 \end{equation}
 and where
 \begin{eqnarray}\label{def-G}
 G_1(t,z,\gamma)
 :=
 \sup_{(v,a)\in\mathcal V\times\mathcal A} \;g_1(t,z,\gamma,v,a),
 \end{eqnarray}
with
 \begin{eqnarray*}
 g_1(t,z,\gamma,v,a)
 &:=&
 -k(v,a)
 +z^\obs\cdot\mu_t(v,a)
 +\frac12\mbox{Tr}\big[\gamma^\obs\Sigma_t(v)\Sigma_t^T(v)\big]
 +\frac12\mbox{Tr}\big[\gamma^\nobs\big] .
  \end{eqnarray*}
Substituting the above into (\ref{eq:eq}), it follows that, $\mathbb P_0-a.s.,$
\begin{align}\label{dyn-VA}
\nonumber d(V^{A,v,a}_te^{R_A K^{v,a}_{0,t}})
 =&
 -R_AV^{A,v,a}_te^{R_A K^{v,a}_{0,t}}\left(g_1(t,Z_t^{v,a},\Gamma_t^{v,a},v_t,a_t)-G_1(t,Z_t^{v,a},\Gamma^{v,a}_t)\right)dt\\
         & -R_AV^A_te^{R_A K^{v,a}_{0,t}}\left(\Sigma_t(v_t)^T Z^{\obs,v,a}_t
                 + \left(\Sigma^\perp_t\right)^T(v_t) Z^{\nobs,v,a}_t
          \right)\cdot dB_t.
 \end{align}
We then see by directly solving the latter stochastic differential equation that
 \begin{align*}
 V^{A,v,a}_T
 =&\
 V_0^{A}\exp\left[R_A\left(\int_0^T\left(G_1(t,Z_t^{v,a},\Gamma_t^{v,a})-\frac12\mbox{Tr}\big[\Gamma^{\obs,v,a}_t
                                                            d\langle B^{\obs,v,a}\rangle_t  \big]-\frac12\mbox{Tr}\big[\Gamma^{\nobs,v,a}_t\big]
                 \right)dt\right)\right]
  \\
  &
  \times \exp\left[-R_A\left(\int_0^T Z^{\obs,v,a}_t\cdot dB^{\obs,v,a}_t
  +\frac{R_A}{2}\int_0^T Z^{\obs,v,a}_t
                                    \cdot
                                    d\langle B^{\obs,v,a}\rangle_t
                                    Z^{\obs,v,a}_t\right)\right]
  \\
&
\times \exp\left[-R_A\left( \int_0^T Z^{\nobs,v,a}_t\cdot dB^{\nobs,v,a}_t
 +\frac{R_A}{2}\int_0^T \No{Z^{\nobs,v,a}_t}^2 dt\right)\right],\ \mathbb P_0-a.s.
 \end{align*}
 Next, we recall that the principal must offer a contract based on the information set $\mathbb{F}^\obs$ only.
 From the definition of $G_1$ we can check that the expression for $\xi_T$  does not depend on $\Gamma^{\nobs,v,a}$, and  we expect
 it also not to depend on $Z^{\nobs,v,a}$, that is, to have $Z^{\nobs,v,a}\equiv 0$. In that case, from \eqref{xiV},
denoting
 \begin{align}
 G_1^\obs(t,z^\obs,\gamma^\obs)
 :=
 \sup_{(v,a)\in\mathcal V\times\mathcal A} \;g^\obs_1(t,z^\obs,\gamma^\obs,v,a),
 \end{align}
and
 \begin{align*}
 g_1^\obs(t,z^\obs,\gamma^\obs,v,a)
 :=
 -k(v,a)
 +z^\obs\cdot\mu_t(v,a)
 +\frac12\mbox{Tr}\big[\gamma^\obs\Sigma_t(v)\Sigma_t^T(v)\big].
 \end{align*}
the contract payoff $\xi_T$ would be as in the following definition:

\begin{Definition}\label{admissible}


 An admissible  contract payoff $\xi_T=\xi_T(Z,\Gamma)$ is a ${\mathcal F}^\obs_T-$measurable random variable that satisfies
\begin{align}
\label{form2}
U_A\left(\xi_T(Z,\Gamma)\right)
 =
 Ce^{-R_A\int_0^T \left\{Z_t\cdot dB^{\obs,v,a}_t
-G^\obs_1(t,Z_t,\Gamma_t)dt
 +\frac12\mbox{\rm Tr}\big[\big(\Gamma_t +R_AZ_tZ^T_t\big)
                                                            d\langle B^{\obs,v,a}\rangle_t
                                                     \big]\right\}}.
\end{align}
for some constant $C<0$, and some  pair $(Z,\Gamma)$ of bounded $\mathbb F^\obs-$predictable processes with values in $\mathbb{R}^2$ and $\mathbb{S}^2$, respectively,
that are such that
there is a maximizer $(v^*(Z,\Gamma),a^*(Z,\Gamma))\in \mathcal U$ of $g^\obs_1(\cdot,Z,\Gamma)$, $dt\times d\mathbb P_0$-a.e..

\vspace{0.3em}
\noindent
We denote by $\mathfrak C
$ the set of all admissible contracts, and by $\mathfrak U
$ the set of the corresponding $(Z,\Gamma)$.
\end{Definition}

\noindent The assumption of boundedness is technical, assumed to simplify the proofs, and it can be relaxed. If, as in the example section,  the optimal $Z$ and $\Gamma$ are constant processes, then the assumption is satisfied. The  assumption that $g^\obs_1(\cdot,Z,\Gamma)$ has a maximizer is needed  to prove the incentive compatibility of  contract $\xi_T$, and to solve the principal's problem.

\begin{Remark}
In addition to a constant term and the $``dt"$ integral term,
 with $U_A$ a CARA utility function, an admissible contract is  is linear $($in the integration sense$)$ in the following factors:
the contractible variables, that is, the output and the contractible sources of risk;
and the quadratic variation and cross-variation processes of the contractible variables.
As seen in the numerical example, the optimal contract generally makes use of
all of these components. This is to be contrasted with
the first best contract,
 and with
the case of controlling the drift only as in
Holmstrom-Milgrom $(1987)$, in which only the output is used,.
\end{Remark}



\begin{Remark} We argue now that, under technical
  conditions, an ``option-like"  contract of the form $\xi_T=F(X^{v,a}_T,B^1_T),$ for a given function $F$ is an admissible contract.
  For notational simplicity, assume $b=0$ and $a=0$,
   and  consider the PDE, with subscripts denoting partial derivatives, and
    with $G_1=G_1(z_1,z_2,\gamma_1,\gamma_2,\gamma_3)$ where $\gamma_1$ and $\gamma_2$ are the diagonal entries of a symmetric  matrix $\gamma$
    and $\gamma_3$ is the value of the off-diagonal entries,
$$u_t
+G_1(u_x,u_y,u_{xx}-R_Au_x^2,u_{yy}-R_Au_y^2,u_{xy}-R_Au_xu_y)=0,\ (t,x)\in[0,T)\times\mathbb R,\ u(T,x,y)=F(x,y).$$
Then, assuming that the  PDE has a smooth solution,
 it follows from It\^o's formula applied to $u(t, X^{v,¿}_t,B^{1}_t)$ that
\begin{align*}
&F(X^{v,0}_T, B^{1}_T)
= u(0,0,0)+\int_0^Tu_x(s, X^{v,0}_s,B^{1}_s)
dX^{v,0}_s
+\int_0^T u_y(s,X^{v,0}_s,B^{1}_s)d B^{1}_s
\\
&
+\int_0^T
\frac12\left( u_{xx}(s, X^{v,0}_s, B^{1}_s)
d\langle X^{v,0}\rangle_s
+u_{yy}(s,X^{v,0}_s, B^{1}_s)d\langle  B^{1}\rangle_s
\right)+\int_0^Tu_{xy}(s, X^{v,0}_s,B^{1}_s)d\langle X^{v,0},B^{1}\rangle_s
\\
&
-\int_0^TG_1^\obs\left(u_x,u_y,u_{xx}-R_Au_x^2,u_{yy}-R_Au_y^2,u_{xy}-R_Au_xu_y\right)(s,X^{v,0}_s, B^{1}_s)ds.
\end{align*}
Thus, $F(X^{v,0}_T, B^{1}_T)$ is of the form $\xi_T(Z,\Gamma)$, where
the vector  $Z_t$ has entries given by $u_x(t,X^{v,0}_t,B^{1}_t)$ and $u_y(t,X^{v,0}_t,B^{1}_t)$,
and where the matrix $\Gamma_t$ has diagonal entries given by $(u_{xx}-R_Au^2_x)(t,X^{v,0}_t, B^{1}_t)$, 
$(u_{yy}-R_Au^2_{y})(t,X^{v,0}_t,B^{1}_t)$,
 and off-diagonal entries given by $(u_{xy}-R_Au_xu_y)(t,X^{v,0}_t,B^{1}_t)$.
\end{Remark}

\noindent At the end  of this section, we show that, as desired, under the above definition of admissible contracts,
one can  characterize the agent's optimal action.
 Introduce the set of  the  controls that are optimal for maximizing $g_1^{obs}$, given $Z,\Gamma$:
 $${\mathfrak V_1}(Z,\Gamma)=\{(v^*(Z,\Gamma),a^*(Z,\Gamma)), ~\hbox{such that the conditions of  Definition \ref{admissible} are satisfied} \}.$$
 The next proposition states that for  a given contract $\xi_T(Z,\Gamma)\in\mathfrak C$,  any  control $(v^*(Z,\Gamma),a^*(Z,\Gamma))\in {\mathfrak V_1}(Z,\Gamma)$ is optimal for the agent.

\begin{Proposition}\label{Prop1_2}
 An admissible contract $\xi_T(Z,\Gamma)\in\mathfrak C$
as defined in \reff{form2} is incentive compatible with $ {\mathfrak V_1}(Z,\Gamma)$. That is, given the contract $\xi_T(Z,\Gamma)$,
any  control $(v^*(Z,\Gamma),a^*(Z,\Gamma))\in {\mathfrak V_1}(Z,\Gamma)$ is optimal for the agent.
Moreover, the corresponding agent's value function satisfies equation \reff{dyn-VA} with $(Z^\obs,Z^\nobs)=(Z,0)$, and $(\Gamma^\obs,\Gamma^\nobs)=(\Gamma,0)$.
\end{Proposition}

\vspace{0.8em}
\noindent {\bf Proof:} Let $(Z,\Gamma)$ be an arbitrary pair process in $\mathfrak U$, and consider the agent's problem with contract $\xi_T(Z,\Gamma)$:
 \begin{eqnarray*}
 V^{A,v,a}_t\big(\xi_T(Z,\Gamma)\big)
  :=
  \underset{(v',a')\in\mathcal U(t,(v,a))}{{\rm{essup}}}
      \mathbb E_t\left[U_A\big(\xi_T(Z,\Gamma)-K_{t,T}^{v',a'})\right],\ \mathbb P_0-a.s.
 \end{eqnarray*}
We first compute, for all $(v',a')\in \mathcal U(t,(v,a))$,
 \begin{align*}
 U_A\left(\xi_T(Z,\Gamma)\right)e^{R_AK^{v',a'}_{t,T}}=&\ U_A\left(\xi_t(Z,\Gamma)\right)\mathcal{E}\Big(-R_A\int_t^.Z_r\cdot \Sigma_r(v'_r)dB_r\Big)_T\\
 &\times \exp\left(
 R_A\int_t^T\big[G_1^\obs(r,Z_r,\Gamma_r)-g_1^\obs(r,Z_r,\Gamma_r, v'_r,a'_r)\big]dr\right)
          ,\ \mathbb P_0-a.s.,
 \end{align*}
where $U_A(\xi_t(Z,\Gamma))$ has the same form \eqref{form2} as $U_A(\xi_T(Z,\Gamma))$, when we substitute $t$ for $T$.

\vspace{0.5em}
\noindent Since $Z$ is bounded by definition and $\sigma$ satisfies the linear growth condition \eqref{eq:lingrowth}, we have by definition of $\mathcal U$ (see \eqref{eq:expmom} in particular) that
$$\mathbb E\left[\exp\left(á\frac{R_A^2}{2}\int_0^T\No{\Sigma_r^T(v'_r)Z_r}^2dr\right)\right]<+\infty.$$
Hence, by Novikov criterion, we may define a probability measure $\bar{\mathbb{P}}$ equivalent to $\mathbb{P}_0$ via the density
$$\frac{d\bar{\mathbb{P}}}{d{\mathbb{P}_0}}|_{{\cal F}_T}=\mathcal{E}\left(-R_A\int_t^.Z_r\cdot \Sigma_r(v'_r)dB_r\right)_T.$$
Then,
 \begin{equation*}
 \mathbb E_t\left[U_A\big(\xi_T(Z,\Gamma)\big)e^{R_AK^{v',a'}_{t,T}}\right]
 =
 U_A\big(\xi_t(Z,\Gamma)\big) \mathbb E^{\bar{\mathbb P}}_t\Big[e^{R_A\int_t^T G_1^\obs(r,Z_r,\Gamma_r)
                                                                             -g_1^\obs(r,Z_r,\Gamma_r,v'_r,a'_r)dr}
                                    \Big].
 \end{equation*}
Since $G_1^\obs-g_1^\obs\ge 0$, we see that
$$\mathbb E_t\left[U_A\big(\xi_T(Z,\Gamma)\big)e^{R_AK^{v',a'}_{t,T}}\right]\le U_A\big(\xi_t(Z,\Gamma)\big),$$
and by the arbitrariness of $(v',a')\in \mathcal U(t,(v,a))$, it follows that $V^{A,v,a}_t(\xi(Z,\Gamma))\le U_A\big(\xi_t(Z,\Gamma)\big)$.

\vspace{0.5em}
\noindent Thus, any control $(v^*,a^{*})$ for  which $G_1^\obs(r,Z_r,\Gamma_r)=g_1^\obs(Z_r,\Gamma_r,v^*_r,a^*_r)$, attains the upper bound. Hence,
 \begin{equation*}
 V^{A,v,a}_t\big(\xi_T(Z,\Gamma)\big)
 =
U_A\big(\xi_t(Z,\Gamma)\big),
 \end{equation*}
and the dynamics of $V_t^{A,v,a}$ are as stated.
\ep

\begin{Remark}
When $U_A$ is not a CARA utility function, the same approach would work if the agent draws utility/disutility
of the form $  U_A(\xi_T)-K^{v,a}_{T}$, that is, if the cost  of effort is separable from the agent's utility function
  \footnote{A possible interpretation is that the function $k$ represents, in a stylized way,
   joint effects of the risk aversion to the choice of $v$ and $a$ and of their cost.}.
   For example, in the case in which
 only $X=X^{v,a}$ is contractible and $\sigma=I_d$, we would define admissible contracts $\xi_T=\xi_T(Z,\Gamma)$ to be those that satisfy
 $$U_A(\xi_T(Z,\Gamma))=
\tilde C+\int_0^TZ_udX_u+\int_0^T\frac{1}{2}\Gamma_u
d\langle X\rangle_u
-\int_0^T\widetilde G_0(Z_u,\Gamma_u) du,
$$
for some constant $\tilde C$,  some $\mathbb F^{X^{v,a}}-$predictable processes $Z$ and $\Gamma$ with values in $\mathbb{R}$, and a process
process $\widetilde G_0(Z_t,\Gamma_t)$ defined similarly as above.
Thus, $U_A(\xi_T)$, rather than $\xi_T$,   would be required to be linear $($in the integration sense$)$. However,
while the agent's problem would be tractable, the difficulty is that, in general, it would be hard to solve the principal's maximization problem.
\end{Remark}

\subsubsection{How general is  class $\mathfrak C$?}\label{sec.discussion}

This question is related to the possibility of  HJB characterization of the value function in the non-Markovian case, which has been approached
  by introducing and studying so-called second-order BSDEs; see, e.g., Soner, Touzi and Zhang (2012).  However,  identification (or even existence) of the optimal control is a very hard task, which may require strong regularity assumptions.
Using results of that recent theory, we prove in  Appendix that, in the case of drift control only (as in Hormstrom and Milgrom (1987)) any feasible contract payoff  has the form \eqref{form2}, while, with volatility control, this is true under additional regularity assumptions. More precisely, we have the following theorem.

\begin{Theorem}\label{th:reg}
If $v$ is uncontrolled and fixed to   $v_0\in\mathcal V$, and if  there is a constant $C>0$ and  $\varepsilon\in[1,+\infty)$ such that
$$\underset{\No{a}\rightarrow +\infty}{\overline{\lim}}\ \frac{k(v_0,a)}{\No{a}}=+\infty,\ \No{D_ak(v_0,a)}\leq C\left(1+\No{a}^\varepsilon\right),$$
$($a condition satisfied in the case of  quadratic cost$)$, then any $\mathcal F_T^\obs$-measurable random variable  $\xi_T$ for which the agent's value function is well-defined can be represented as in \eqref{form2}.
  More generally, \eqref{form2} is implied by ``smoothness" of the non-martingale component of the agent's value function in the sense of   existence of a ``second-order sensitivity"  process $\Gamma$ as in  Assumption \ref{assump:reg} in Appendix.
\end{Theorem}

\noindent Basically, by imposing the form \eqref{form2} with an appropriately chosen $G_1^\obs$, we  ensure that the corresponding contract payoff is sufficiently smooth to  make the agent's value function process also smooth. In other words, with such a contract, the value function of the agent is a  solution to the path-dependent HJB equation that we derived heuristically in \eqref{eq:HJB}. In the existing literature (in which only drift
control is present)  this is always the case, whether the model is with finite or infinite horizon. Indeed, the martingale representation and the comparison theorem type result is always used, which is equivalent to the path dependent HJB equation. It seems unlikely that one could solve the agent's problem for contracts for which the value function would be so degenerate that it would not satisfy this weak form of smoothness.   In practice, this means that the principal also wouldn't know what the agent would do  given such contracts, and the principal would likely
decide not to consider them.

\subsection{Attainability of first best.}

 In this section we assume that drift effort $a$ is uncontrolled by the agent and fixed to the value of zero. Moreover, we adopt the setting of \eqref{eq:simple}, with $\mathcal V=\mathbb R^d$ and CARA utility functions.
 We  show that first best is always attainable when the cost function is zero. In other words,
 with no constraints on the choice of $v$ and no cost of choosing it, there is no agency friction.

\vspace{0.5em}
\noindent For  a pair  $(Z,\Gamma)\in\mathbb R^2\times\mathbb S_2$, we introduce the notation
$$Z:=\begin{pmatrix}
z_1\\
z_2
\end{pmatrix}
\text{ and }\Gamma:=\begin{pmatrix}
\gamma_1& \gamma_2\\
\gamma_2 &\gamma_3
\end{pmatrix}.$$
If we choose $\gamma_1< 0$ and assume that the cost $k(v,0)$ is $0$, then it is easy to see that the optimal $v^*\in\mathbb R^d$ in the definition of $G_1^\obs(Z,\Gamma)$ is the unique solution of the following equation
$$z_1\sigma b+\gamma_2\sigma_{.1}+\gamma_1\sum_{i=1}^d\sigma_{.i}\cdot v^*\sigma_{.i}=0,$$
where for any $1\leq i\leq d$, $\sigma_{.i}$ is the $i$-th column of matrix $\sigma$. This leads directly to
 \begin{eqnarray}\label{nu*remark}
 v^*
 =
 -\frac1\gamma_1(\sigma\sigma^T)^{-1}\left(z_1\sigma b+\gamma_2\sigma_{.1}\right).
 \end{eqnarray}
The principal may choose the values
 \beaa
 z_1:=\frac{R_P}{R_A+R_P},
 ~~\gamma_1:=-\frac{R_AR_P^2}{(R_A+R_P)^2},
 &\gamma_2:=0,&
 \eeaa
and arbitrary values for $z_2$ and $\gamma_3$. Then, this contract is clearly admissible, since $Z$ and $\Gamma$ are bounded and constant,
  and it follows from \eqref{nu*remark} that when $\sigma=I_d$, it implements the following optimal volatility
$$
v^*=\frac{R_A+R_P}{R_AR_P}b.
$$
When $k=0$, this is exactly equal to the first best volatility $v^{FB}$ obtained in \reff{firstbest} (for $d=2$, but it holds for any $d$), as can easily be verified, and
the same holds for general $\sigma$.
This result is in agreement with Cadenillas, Cvitani\'c and Zapatero (2007),
who show that in a frictionless and a complete market  (i.e., zero cost of volatility effort and the  number of Brownian motions equal to the number of observable assets), with arbitrary utility functions,   first best is attainable using a contract
that depends only on the final value of the output.
We see here that with exponential utility functions completeness is not necessary, and a linear contract is optimal. Indeed, with the above $z_1$ and $\Gamma$, we have, setting $z_2=0$,
$$\xi_T(Z,\Gamma)=const. + \frac{R_P}{R_A+R_P}X_T,$$
which is the same as the first best contract when $k(v)=0$.



\subsection{The principal's problem with CARA utility}

In this section, we assume that the utilities are exponential for both the principal and the agent, that is we have
$U_I(x)=-e^{-R_Ix}$, $I=R_A,R_P$.
Fix
 an admissible $\xi_T(Z,\Gamma)\in\mathfrak C
 $ and introduce the notation
$$Z=\begin{pmatrix}
Z^X\\
Z^{B^1}
\end{pmatrix}
\text{ and }\Gamma=\begin{pmatrix}
\Gamma^X& \Gamma^{XB^1}\\
\Gamma^{XB^1} &\Gamma^{B^1}
\end{pmatrix}.$$
The principal maximizes the expected utility of her terminal payoff $X^{v,a}_T-\xi_T(Z,\Gamma)$. Since the contract $\xi_T(Z,\Gamma)$ is incentive compatible in the sense of Proposition \ref{Prop1_2}, the optimal volatility and drift choices by the agent correspond to any $(v^*(Z,\Gamma),a^*(Z,\Gamma))\in {\mathfrak V_1}(Z,\Gamma)$. Then, assuming that the agent lets the principal choose among the control choices that are optimal for the agent,
 the principal problem is:
 $$
 \sup_{(Z,\Gamma )\in\mathfrak U~, ~(v^*(Z,\Gamma),a^*(Z,\Gamma))\in {\mathfrak V_1}(Z,\Gamma)} \
 \mathbb E\big[U_P\big(X^{v,a}_T- \xi_T(Z,\Gamma )\big)\big].
 $$
Denoting $(v^*,a^*):=(v^*(Z,\Gamma),a^*(Z,\Gamma))$, and substituting the expression for $\xi_T(Z,\Gamma)$, we get
 \begin{align*}
 X^{v,a}_T- \xi_T(Z,\Gamma )
 =&\ -C+
 \int_0^T\Big(\sigma_s^T(v^*_s)b_s(a^*_s)(1-Z_s^X)
                     -\frac{1}{2}\big\|\sigma_s(v_s ^*)\big\|^2
                       \big(\Gamma_s^X+R_A\big|Z_s^X\big|^2\big)
              \Big) ds
 \\
 &
 +\int_0^T\Big(G_1^\obs(s,Z_s,\Gamma_s)
                       -\frac12\big(\Gamma_s^{B^1}+R_A\big|Z_s^{B^1}\big|^2\big)
               \Big)ds
\\
&-\int_0^T\big(\Gamma^{XB^1}_s+R_AZ_s^XZ_s^{B^1}\big)d\langle X, B^{1}\rangle_s
\\
&
-\int_0^T Z^{B^1}_sd(B^{\obs,v^*,a^*})^{1}_s+\int_0^T\big(1-Z^X_s\big) v_s ^*\cdot \sigma d B^{v^*,a^*}_s,\ \mathbb{P}_0-\mbox{a.s.}
 \end{align*}
  Introduce a vector $\theta^*:=\sigma_s(v^*_s)$ and denote its first entry $\theta^*_1(s)$, and denote  by $\theta^*_{-1}(s)$ the $(d-1)$-dimensional vector without the first entry. Arguing exactly as in the proof of Proposition \ref{Prop1_2}, in particular, by isolating the appropriate stochastic exponential, it follows that the principal problem reduces to maximizing
 \begin{align}\label{finalopt}
 \nonumber  \theta_s^*(&Z,\Gamma) \cdot b_s(a^*_s(Z,\Gamma))
\left(1-Z^X
\right)-\frac{1}{2}\No{ \theta_s^*(Z,\Gamma)}^2
\big(\Gamma^X+R_A\big|Z^X\big|^2\big)
+ G_1^\obs(s,Z,\Gamma)\\
&-\frac{1}{2}\big(\Gamma^{B^1}+R_A\big|Z^{B^1}\big|^2\big)-
\big(\Gamma^{XB^1}+R_AZ^XZ^{B^1}\big)\theta_1(s,Z,\Gamma)
\nonumber
\\  & -\frac{R_P}{2}\left[\No{\theta^*_{-1}(s,Z,\Gamma)}^2\big(1-
Z^X\big)^2  +\big(\theta^*_{1}(s,Z,\Gamma)\big(1-
Z^X\big)-
Z^{B^1}\big)^2\right].
\end{align}
Since the supremum in the definition of $G_1^\obs(s,Z,\Gamma)$ is attained at $(v^*,a^*)$, this is equivalent to the minimization problem
 \begin{align}
 \nonumber
&\inf_{Z,\Gamma,v^*
(Z,\Gamma),a^*(Z,\Gamma)} \Big\{ -
\theta^*(s,Z,\Gamma)\cdot b_s(a^*_s(Z,\Gamma))
+\frac{1}{2}\No{  \theta^*(Z,\Gamma)}^2R_A\left(Z^X\right)^2
+
k(v^*,a^*)+\frac{R_A}{2}\left(Z^{B^1}\right)^2\\
 &+
R_AZ^XZ^{B^1}\theta^*_1(s,Z,\Gamma)
+
\frac{R_P}{2}\Big[\No{\theta^*_s(Z,\Gamma)}^2\left(1-Z^X\right)^2
 -2\theta^*_{1}(s,Z,\Gamma)\left(1-Z^X\right)Z^{B^1}+
 \left(Z^{B^1}\right)^2\Big]
 \Big\}.
 \label{finalopt2}
\end{align}
Note that if minimizers $Z^*$, $\Gamma^*$, $v^*$ and $a^*$ exist,
they are then necessarily deterministic,
since $b$, $\sigma$ and $k$ are non-random.
By Proposition \ref{Prop1_2}, the contract $\xi_T(Z^*,\Gamma^*)$ is incentive compatible for $(v^*(Z^*,\Gamma^*),a^*(Z^*,\Gamma^*))$,
if $\xi_T(Z^*,\Gamma^*)\in\mathfrak C$.
Moreover, as we have just shown, it is also optimal for the principal's problem, that is, we have proved the following.

\begin{Theorem}
Consider the set of admissible contracts $\xi_T(Z,\Gamma)\in\mathfrak C
$. Then, the  contract that is optimal in that set and
 provides the agent with expected utility $V_0^A<0$ is $\xi_T(Z^*,\Gamma^*)$ corresponding to $Z^*,\Gamma^*,v^*,a^*$ which are the minimizers in \reff{finalopt2}, provided such minimizers exist, with $v^*\neq 0$.
  Moreover,
 the contract cash constant $C$ is given by
$C:=-\frac{1}{R_A}\log(-V_0^A)$.
\end{Theorem}

\noindent {\bf Proof.}
The only thing to check here is the admissibility of the contract $\xi_T(Z^*,\Gamma^*)$, but this is just a consequence of the optimizers being deterministic, noting
 that  condition \eqref{eq:invertible} is  satisfied when $v^*\neq 0$.\ep


\vspace{0.5em}
\noindent
Next, we provide  here  sufficient conditions for existence of at least one minimizer of \eqref{finalopt2} when, for simplicity, there is no optimization with respect to $a$, and when the cost function $k$ is super-quadratic. The case of a quadratic $k$ is actually harder and  is treated in Appendix  in Proposition \ref{prop:max}.

\begin{Proposition}\label{prop.growth}
Assume that the agent does not control the drift, i.e. $a=0$, and consider the setting of \eqref{eq:simple} with $\mathcal V=\mathbb R^d$. Assume moreover that the cost function $k(v):=k(v,0)$ is at least $C^1$ and satisfies for some constant $C>0$
$$\|\nabla k(v)\|\leq C\left(1+\|v\|^{1+\varepsilon}\right),\ \text{for some $\varepsilon >0$, and }\underset{\|v\|\rightarrow +\infty}{\underline{\lim}}\ \frac{k(v)}{\|v\|^2}=+\infty.$$
Then, the infimum in \eqref{finalopt2} is attained.
\end{Proposition}

\section{Second-best with non-contractible risks}

Consider now the case in which the only contractible process is  $X^{v,a}$.
In that case, we need to modify
our approach by adopting the following changes, as can be verified using similar arguments. First of all, the principal can now only offer contract payoffs measurable with respect to ${\cal F}^{X^{v,a}}_T$, a sigma-field contained in the filtration $\mathbb F^{X^{v,a}}:=\{{\cal F}^{X^{v,a}}_t\}_{0\leq t\leq T}$ generated by the output process $X^{v,a}$.

\vspace{0.5em}
\noindent Following exactly the same intuition from the stochastic control theory as in the previous section, we introduce the function $G_0^\obs$, the counterpart of the function $G_1^\obs$ above, defined for any $(s,z,\gamma)\in[0,T]\times\mathbb R\times\mathbb R$ by
 \begin{align*}
 G_0^\obs(s,z,\gamma)
 &:= \sup_{ (v,a)\in\mathcal V\times\mathcal A }g_0^\obs(s,z,\gamma,v,a)
 \\
 &:=
 \sup_{ (v,a)\in\mathcal V\times\mathcal A }\Big\{\sigma_s^T(v)b_s(a)  z
+\frac{1}{2} \No{\sigma_s (v)}^2\gamma
-
k( v,a )
\Big\},
~~z,\gamma\in\mathbb{R}.
 \end{align*}
Once again, if a maximizer exists, we denote it by $ (v ^*(z,\gamma),a^*(z,\gamma))$. We now introduce the set of admissible contracts in this case.

\begin{Definition}\label{admissible2}
An admissible  contract payoff $\xi_T=\xi_T(Z,\Gamma)$ is  an ${\mathcal F}^{X^{v,a}}_T-$measurable random variable that satisfies
\begin{align}
\label{form3}
U_A\left(\xi_T^0(Z,\Gamma)\right)
 :=
 Ce^{-R_A\int_0^T \left\{Z_t dX^{v,a}_t
-G^\obs_0(t,Z_t,\Gamma_t)dt
 +\frac12\big(\Gamma_t +R_AZ_t^2\big)
                                                            d\langle X^{v,a}\rangle_t
                                                     \big]\right\}}.
\end{align}
for some constant $C\geq0$, and some  pair $(Z,\Gamma)$ of bounded $\mathbb F^{X^{v,a}}-$predictable processes with values in $\mathbb{R}$,
%
 and such that there is a maximizer $(v^*(Z,\Gamma),a^*(Z,\Gamma))\in \mathcal U$ of $g^\obs_0(\cdot,Z,\Gamma)$, $dt\times d\mathbb P_0$-a.e..

\vspace{0.3em}
\noindent
We denote by $\mathfrak C_0
$ the set of all admissible contracts, and by $\mathfrak U_0
$ the set of the corresponding $(Z,\Gamma)$.
\end{Definition}

\noindent Similarly as before, we introduce the set of controls that are optimal for maximizing $g_0^\obs$, given $Z,\Gamma$:
 $${\mathfrak V}_0(Z,\Gamma)=\{(v^*(Z,\Gamma),a^*(Z,\Gamma)), ~\hbox{such that the conditions of  Definition \ref{admissible2} are satisfied} \}.$$
The following proposition is the analogue of Proposition \ref{Prop1_2} in this setting, and can be proved by the same argument.
\begin{Proposition}\label{Prop1_2B}
An admissible contract $\xi_T^0(Z,\Gamma)
$, as defined in \reff{form3}, is incentive compatible with $ {\mathfrak V}_0(Z,\Gamma)$. That is, given the contract $\xi^0_T(Z,\Gamma)$,
 any control in $\mathfrak V_0(Z,\Gamma)$ is optimal for the agent's problem.
\end{Proposition}

\noindent Accordingly, the principal's problem is modified as follows, denoting again for notational simplicity $ (v^*,a^*):= (v^*(Z,\Gamma),a^*(Z,\Gamma))$, and assuming $U_P(x)=-e^{-R_Px}$:
\begin{align*}
\inf_
{(Z,\Gamma,v^*,a^*)\in{\mathfrak U}_0\times{\mathfrak V}_0(Z,\Gamma) }\
\mathbb E\bigg[&\exp\bigg\{R_P\bigg(\int_0^T\left(\sigma_s^T(v_s^*)b_s(a^*_s) (Z_s-1
)
+\frac{1}{2}\No{\sigma_s(v_s ^*)}^2
\left(\Gamma_s+R_AZ_s^2\right)
\right)ds\\
&-\int_0^T
  G_0^\obs(u,Z_u,\Gamma_u)du+
\int_0^T (
Z_s-1)  dX^{v^*,a^*}_s\bigg)
\bigg\}\bigg].
\end{align*}
Similarly, as above, denote by $ \theta^*_s$ the vector $\sigma_s(v^*_s)$.
The principal's problem then becomes
\begin{align}\label{finaloptB}
&
\inf_{Z,\Gamma,
 v^*,a^*} \Big\{-\theta^*_s(Z,\Gamma)b_s(a^*_s)
+\frac{R_A}{2}\No{   \theta_s^*(Z,\Gamma)}^2Z^2
+k\left( v^*,a^*\right)
 +\frac{
R_P}{2}\No{\theta^*_s(Z,\Gamma)}^2(1
-
Z)^2
\Big\}.
\end{align}
We then have, similarly as before,
\begin{Theorem}\label{th:noncontract}
Consider the set of admissible contracts $\xi_T^0(Z,\Gamma)\in\mathfrak C_0$. Then, the optimal contract that provides the agent with optimal expected utility $V_0^A$ is the one corresponding to  $Z^*,\Gamma^*,v^*,a^*$ which are the minimizers in \reff{finaloptB},
  provided such minimizers exist with $v^*\neq 0$.
    Moreover, the contract cash term in the contract is given by
$-\frac{1}{R_A}\log(-V_0^A)$.
\end{Theorem}

\noindent The analogue of  Proposition \ref{prop.growth} still holds in this context, with the  same  statement and the same  proof.

\section{Conclusions}
\setcounter{equation}{0} \setcounter{Assumption}{0}
\setcounter{Theorem}{0} \setcounter{Proposition}{0}
\setcounter{Corollary}{0} \setcounter{Lemma}{0}
\setcounter{Remark}{0}
\setcounter{Definition}{0}\setcounter{Example}{0}
We build a framework for studying moral hazard in dynamic risk management,
using recently developed mathematical techniques. While those allow us to solve
the problem in which utility is drawn solely from terminal payoff,
we leave for a future paper the similar problem on infinite horizon  with inter-temporal payments, a la Sannikov (2008).
In the case of terminal payoff, we find  that the optimal contract
is implemented
by compensation based on the output, its quadratic variation (corresponding in practice to the sample variance used when
computing Sharpe ratios), the contractible sources of risk,
and the cross-variations between the output and the risk sources.
(Or, it could be implemented
by derivatives that provide such payoffs.)
It is an open question how much, if any, mathematical generality we lose in restricting the family of admissible contracts the way we do.
We argue that from a practical perspective, we do include  all the contracts that are likely to be considered by a principal  in the
real world.
In our framework it is assumed that the principal knows the model parameters (for example, the mean return rates and
the variance-covariance matrix of the returns of the assets the hedge fund manager is investing in). In practice,  the principal may not have that information, and it would be of interest to extend the model to include this adverse selection
problem. This might be approached by combining our model with the one  of Leung (2014).


\section{Appendix}
\setcounter{equation}{0} \setcounter{Assumption}{0}
\setcounter{Theorem}{0} \setcounter{Proposition}{0}
\setcounter{Corollary}{0} \setcounter{Lemma}{0}
\setcounter{Remark}{0}
\setcounter{Definition}{0}\setcounter{Example}{0}

 {\bf Proof of Theorem \ref{th:reg}.} We  adapt  arguments of Soner, Touzi \& Zhang (2011, 2012, 2013). In Step $1$ we transform our problem to
  the weak formulation used in those papers. The agent's problem is then analyzed in Step $2$. Finally, Step $3$ specializes to the case of uncontrolled volatility.

 \vspace{0.5em}
 \noindent {\bf Step $1$: An alternative formulation of the agent's problem}

 \vspace{0.5em}
 \noindent Let us  consider the following family of processes, indexed by admissible processes $v$:

\begin{equation}\label{Xv}
M^{v}_\cdot:=\begin{pmatrix}\displaystyle \int_0^\cdot \sigma_s(v_s) \cdot dB_s
\\
B^1_\cdot\\
\vdots\\
B^{d_0}_\cdot\\
\displaystyle \int_0^\cdot\Sigma_s^\perp(v_s)dB_s
\end{pmatrix}=\begin{pmatrix}\displaystyle \int_0^\cdot\Sigma_s(v_s)dB_s
\\
\displaystyle \int_0^\cdot\Sigma_s^\perp(v_s)dB_s
\end{pmatrix},\ \mathbb P_0-a.s.,
\end{equation}
where, similarly as in Section \ref{sec:sec}, for any $s\in[0,T]$ and any $v\in\mathcal V$ the $(d_0+1)\times d$ matrix $\Sigma(v)$ is defined by
$$ \Sigma_s(v):=\begin{pmatrix}\displaystyle \sigma_s^T(v)\\
I_{d_0,d}
\end{pmatrix},\text{Êwith } I_{d_0,d}:=\begin{pmatrix}I_{d_0} & 0_{d_0,d-d_0}\end{pmatrix}.$$ Furthermore, $\Sigma^\perp_s$ is now a $(d-d_0-1)\times d$ matrix satisfying, for any $s\in[0,T]$ and any $v\in\mathcal V$,
$$\Sigma^\perp_s(v)\Sigma_s(v)^T=0_{d-d_0-1,d_0+1}\ \text{Êand }\Sigma_s^\perp(v)(\Sigma_s^\perp)^T(v)=I_{d-d_0-1}.$$
We then set $\mathcal P_m$ to be the set of probability measures $\mathbb P^v$ on $(\Omega, \mathcal F)$  of the form
\begin{equation}\label{prob-strong}
\mathbb P^v:=\mathbb P_0\circ\left(M^{v}_.\right)^{-1},\text{Ê for any admissible $v$}.
\end{equation}
We recall that by Bichteler (1981), we can also define a pathwise version of the quadratic variation process $\langle  B\rangle$ and of its  density  process with respect to the Lebesgue measure, a positive symmetric matrix $\widehat \alpha$:
$$\widehat \alpha_t:=\frac{d\langle B\rangle_t}{dt}~.$$
The elements of family
 correspond to possible choices of the volatility vector $\sigma_s(v)$ by the agent.\footnote{This family could also be characterized
  by considering all the choices of control $u$ for which there exists at least one strong solution to the SDE for $M$ (see Soner, Touzi and Zhang (2011)), or, equivalenntly,  to a certain  martingale problem (see Kazi-Tani, Possama\"i and Zhou (2013) and Neufeld and Nutz (2014)).} We remark that by our definitions, we have teh following weak formulation:
$$
\text{The law of $(B,\widehat \alpha\ )$ under $\mathbb P^v$ $=$ the law of $\left(M^v,\begin{pmatrix}
\Sigma_s(v)\Sigma_s(v)^T & 0_{d_0+1,d-d_0-1}\\
0_{d-d_0-1,d_0+1} & I_{d-d_0-1}
\end{pmatrix}\right)$ under $\mathbb P_0$.}$$
Moreover, exactly as in Section \ref{sec:sec}, the density of the quadratic variation of $B$ is invertible under $\mathbb P^v$, if and only if $ \Sigma_s(v)\Sigma_s(v)^T$ is invertible, which can be shown to be equivalent to
$$\No{\sigma_s(v_s)}^2-\sum_{i=1}^{d_0}\abs{\sigma_s^i(v_s)}^2\neq 0,$$
which is the same as \eqref{eq:invertible}.

\vspace{0.5em}
\noindent Then, according to Lemma 2.2 in Soner, Touzi, Zhang (2013), for every admissible $v$, there exists a $\mathbb F$-progressively measurable mapping $\beta_v:[0,T]\times\Omega\longrightarrow \mathbb R^d$ such that
$$B=\beta_v(M^v),\ \mathbb P_0-a.s.,\ \widehat \alpha_s(B)=\begin{pmatrix}
\Sigma_s(\beta_v(B))\Sigma_s^T(\beta_v(B)) & 0_{d_0+1,d-d_0-1}\\
0_{d-d_0-1,d_0+1} & I_{d-d_0-1}
\end{pmatrix},\ \mathbb P^v-a.s.$$
This implies in particular, that the process
\begin{equation}\label{eq:W}
W_t:=\int_0^t\alpha_s^{-1/2} dB_s,\ \mathbb P-a.s,
\end{equation}
is a $\mathbb R^d$-valued, $\mathbb P$-Brownian motion, for every $\mathbb P\in\mathcal P_m$\footnote{Notice that we should actually have considered a family $(W^\mathbb P)_{\mathbb P\in\mathcal P_m}$, since the stochastic integral in \eqref{eq:W} is, a priori, only defined $\mathbb P$-a.s. However, we can use  results of Nutz (2012) to provide an aggregated version of this family, which is the process we denote by $W$. That result  holds under a "good" choice of set theoretic axioms that we do not specify here.}. In particular, this implies that the canonical process $B$ admits the following dynamics, for every admissible $v$,
\begin{align}\label{model}
\nonumber B^1_t&=B^1_0+\int_0^t\sigma_s(v_s(W_\cdot))\cdot dW_s,\ \mathbb P^v-a.s.,\\
\nonumber B^j_t&=B^j_0+W_t^{j-1},\ \mathbb P^v-a.s.,\ j=2\dots d_0+1,\ \text{if $d_0>0$,}\\
\begin{pmatrix}
B^{d_0+2}_t\\
\vdots\\
B^d_t
\end{pmatrix}&=\begin{pmatrix}
B^{d_0+2}_0\\
\vdots\\
B^d_0
\end{pmatrix}+\int_0^t\Sigma_s^\perp\left(v_s(W_\cdot)\right)dW_s,\ \mathbb P^v-a.s.
\end{align}

\vspace{0.5em}
\noindent Thus, the first coordinate of the canonical process is the desired output process, observed by both the principal and the agent, the following $d_0$ coordinates  represent the contractible sources of risk, while the  remaining ones represent the factors that are not contractible.

\vspace{0.5em}
\noindent The introduction of the controlled drift $b_s(a_s)$ can now be done by using  Girsanov transformations
\footnote{ Note that by assumption that the first $d_0$ entries  of vector $b$  do not depend on $a$, the choice of control $a$ does not modify the distribution of the exogenous sources of risk  $(B^2,\dots,B^{d_0+1})$.}.
 We define for any $(v,a)\in\mathcal U$ and any $\mathbb P^v\in \mathcal P_m$, the equivalent probability measure $\mathbb P^{v,a}$ by
\begin{align*}
\frac{d\mathbb P^{v,a}}{d\mathbb P^v}:=\mathcal E\left(\int_0^\cdot b_s(a_s)\cdot dW_s\right)_T,
\end{align*}
and we denote by $\mathcal P:=\left(\mathbb P^{v,a}\right)_{(v,a)\in\mathcal U}$.

\vspace{0.5em}
\noindent Next, by Girsanov theorem, the following process $W^{a}$ is a $\mathbb P^{v,a}$-Brownian motion
$$W^{a}_t:=W_t-\int_0^tb_s(a_s)ds,\ \mathbb P^{v,a}-a.s.\text{ (thus also $\mathbb P^v-a.s.$)}$$

\vspace{0.5em}
 \noindent Then, by \reff{model}, we have
\begin{align}\label{model22}
\nonumber B^1_t&=B^1_0+\int_0^t\sigma_s(v_s(W_\cdot))\cdot \left(b_s(a_s)ds+dW^a_s\right),\ \mathbb P^{v,a}-a.s.,\\
\nonumber B^j_t&=B^j_0+W_t^{j-1},\ \mathbb P^{v,a}-a.s.,\ j=2\dots d_0+1,\\
\begin{pmatrix}
B^{d_0+2}_t\\
\vdots\\
B^d_t
\end{pmatrix}&=\begin{pmatrix}
B^{d_0+2}_0\\
\vdots\\
B^d_0
\end{pmatrix}+\int_0^t\Sigma_s^\perp\left(v_s(W_\cdot)\right)dW_s,\ \mathbb P^{v,a}-a.s.,
\end{align}
which can then be rewritten as
\begin{align}\label{model2}
\nonumber B^{\obs}:=\begin{pmatrix}
B^{1}_t\\
\vdots\\
B^{d_0+1}_t
\end{pmatrix}&=\begin{pmatrix}
B^{1}_0\\
\vdots\\
B^{d_0+1}_0
\end{pmatrix}+\int_0^t\mu_s(v_s(W_\cdot),a_s)ds+\int_0^t\Sigma_s(v_s(W_\cdot))dW^a_s,\ \mathbb P^{v,a}-a.s.,\\
B^\nobs:=\begin{pmatrix}
B^{d_0+2}_t\\
\vdots\\
B^d_t
\end{pmatrix}&=\begin{pmatrix}
B^{d_0+2}_0\\
\vdots\\
B^d_0
\end{pmatrix}+\int_0^t\Sigma_s^\perp\left(v_s(W_\cdot)\right)dW_s,\ \mathbb P^{v,a}-a.s.,
\end{align}
where for any $s\in[0,T]$ and any $(v,a)\in\mathcal U$, $\mu_s(v,a)$ is a $\mathbb R^{d_0+1}$ vector defined by
$$\mu_s(v,a):=\begin{pmatrix}
\sigma_s^T(v)b_s(a)\\
\begin{pmatrix}I_{d_0} & 0_{d_0,d_0-d}
\end{pmatrix}
\end{pmatrix}.$$
Notice then that, for a given measure $\mathbb P\in\mathcal P$, according to \eqref{model2}, we can always find two $m$-dimensional and $n-$dimensional vectors $v^\mathbb P$ and $a^\mathbb P$ such that
$$B^\obs_t=B^\obs_0+\int_0^t\mu_s\left(v_s^\mathbb P,a^\mathbb P_s\right)ds+\int_0^t\Sigma_s\left(v_s^\mathbb P\right)dW^{a^\mathbb P}_s,\ \mathbb P-a.s.,
$$
which gives us the following correspondence
\begin{equation*}
v^{\mathbb P^{v,a}}(B_\cdot)=v(W_\cdot), \ a^{\mathbb P^{v,a}}(B_\cdot)=a(B_\cdot),\ dt\times\mathbb P^{v,a}-a.e.
\end{equation*}
\noindent In particular, this implies that $v^{\mathbb P^{a,v}}=v^{\mathbb P^{v}}$, for any admissible control $a$. We will therefore always write $v^{\mathbb P^v}$ from now on.

\vspace{0.5em}

\noindent Using the above notation, we re-write the  value function of the agent as:
\begin{equation}\label{eq:valueweak}
V^{A,w}_t:=\underset{\mathbb P^{'}\in\mathcal P(t,\mathbb P)}{{\rm essup}^\mathbb P}\ \mathbb E_t^{\mathbb P^{'}}\left[U_A(\xi_T-K_{t,T}^{\mathbb P^{'}})\right],\ \mathbb P-a.s., \ \text{for all }\mathbb P\in\mathcal P,
\end{equation}
where for any $\mathbb P\in\mathcal P$, the set $\mathcal P(t,\mathbb P)$ is the set of probability measures in $\mathcal P$ which agree with $\mathbb P$ on $\mathcal F_t$, and where we have
$$K_{t,T}^\mathbb P:=\int_t^T k(v_s^\mathbb P,a_s^\mathbb P)ds.$$

\vspace{0.5em}
\noindent The definition of the value function depends {\it a priori} explicitly on the measure $\mathbb P$, and we should instead have defined a family $(V_t^{A,w,\mathbb P})_{\mathbb P\in\mathcal P}$. Indeed, it is not immediately clear whether this family can be aggregated into a universal process $V^{A,w}$ or not. Such problems are inherent to the weak formulation of stochastic control problems involving volatility control of the diffusion, see Soner, Touzi and Zhang (2011), Nutz and Soner (2012), Nutz and van Handel (2013), Possama\"i, Royer and Touzi (2013), Epstein and Ji (2013). In our context, it suffices to remark that following similar arguments as in Section 5 of Possama\"i, Royer and Touzi (2013), one can show that  family $\mathcal P$ satisfies their condition $5.4$, and to remark that their approach can be extended to  non-martingale measures (as in Nutz and van Handel (2013)). This allows us to define properly the value function of the agent\footnote{Notice that to be completely rigorous, we should then introduce the so-called universal filtration on $\Omega$, completed by the polar sets generated by $\mathcal P$, to which the process $V^{A,w}$ would then be adapted; see the references mentioned above}.

\vspace{0.5em}
\noindent {\bf Step $2$: Solving the agent's problem}

\noindent Let us start by fixing some $(v,a)\in\mathcal U$. Then, under sufficient integrability conditions for $\xi_T$, the process $(e^{R_AK_{0,t}^{\mathbb P^{v,a}}}V_t^{A,w})_{0\leq t\leq T}$ is a c\`adl\`ag, $\mathbb P^{v,a}$-supermartingale for the filtration $\mathbb F$, which is equal to $\mathbb F^\obs\vee\mathbb F^\nobs$. Using results of Soner, Touzi and Zhang (2012), we know that the martingale representation property still holds under $\mathbb P^{v,a}$, so that, by Doob-Meyer's decomposition, there exists a pair of processes $\tilde Z^{v,a,\obs}$ and $\tilde Z^{v,a,\nobs}$, which are respectively $\overline{\mathbb F^{\obs}}^{\mathbb P^{v,a}}$ and $\overline{\mathbb F^{\nobs}}^{\mathbb P^{v,a}}$-predictable process, and an $\overline{\mathbb F}^{\mathbb P^{v,a}}$-adapted process $\tilde A^{v,a}$, which is non-decreasing $\mathbb P^{v,a}-a.s.$, such that, after applying It\^o's formula, we have the decomposition
\begin{align*}
V_t^{A,w}=& U_A(\xi_T)+\int_t^TR_AV_s^{A,w} k(v_s^{\mathbb P^{v}},a_s^{\mathbb P^{v,a}})ds-\int_t^Te^{-R_AK_{0,s}^{\mathbb P^{v,a}}}\tilde Z^{v,a,\obs}_s\cdot\Sigma_s(v_s^{\mathbb P^{v}})dW^{a^{\mathbb P^{v,a}}}_s\\
&-\int_t^Te^{-R_AK_{0,s}^{\mathbb P^{v,a}}}\tilde Z^{v,a,\nobs}_s\cdot\Sigma_s^\perp(v_s^{\mathbb P^{v}})dW^{a^{\mathbb P^{v,a}}}_s+\int_t^Te^{-R_AK_{0,s}^{\mathbb P^{v,a}}}d\tilde A^{v,a}_s,\ \mathbb P^{v,a}-a.s.
\end{align*}
By definition of $W^{a}$, we deduce that, $\mathbb P^{v,a}-a.s.,$
\begin{align*}
V_t^{A,w}=&U_A(\xi_T)-\int_t^TR_AV_s^{A,w}\Big(\mu_s(v_s^{\mathbb P^{v}},a_s^{\mathbb P^{v,a}})\cdot  Z_s^{v,a,\obs}+\ell_s(v_s^{\mathbb P^{v}},a_s^{\mathbb P^{v,a}})\cdot  Z_s^{v,a,\nobs}-k(v_s^{\mathbb P^{v}},a_s^{\mathbb P^{v,a}})\Big)ds\\
&+\int_t^TR_AV_s^{A,w} Z^{v,a,\obs}_s\cdot\Sigma_s(v_s^{\mathbb P^{v}})dW_s+\int_t^TR_AV_s^{A,w} Z^{v,a,\nobs}_s\cdot\Sigma_s^\perp(v_s^{\mathbb P^{v}})dW_s -\int_t^TR_AV_s^{A,w}dA_s^{v,a},
\end{align*}
where we defined
$$Z^{v,a,\obs}_t:=-\frac{e^{-R_AK_{0,t}^{\mathbb P^{v,a}}}}{R_AV_t^{A,w}}\tilde Z^{v,a,\obs}_t,\ Z^{v,a,\nobs}_t:=-\frac{e^{-R_AK_{0,t}^{\mathbb P^{v,a}}}}{R_AV_t^{A,w}}\tilde Z^{v,a,\nobs}_t,\ A_t^{v,a}:=-\int_0^t\frac{e^{-R_AK_{0,s}^{\mathbb P^{v,a}}}}{R_AV_s^{A,w}}d\tilde A^{v,a}_s,$$
and for any $s\in[0,T]$ and any $(v,a)\in\mathcal V\times\mathcal A$
$$\ell_s(v,a):=\Sigma_s^\perp(v)b_s(a).$$
Next, using the pathwise construction of the quadratic co-variation of Bichteler (1981), it is actually possible to aggregate the families $(Z^{v,a,\obs})_{(v,a)\in\mathcal U}$, $(Z^{v,a,\nobs})_{(v,a)\in\mathcal U}$ into universal processes $Z^\obs$ and $Z^{\nobs}$ (see for instance Nutz \& van Handel (2013)), so that we obtain for all $(v,a)\in\mathcal U$, $ \mathbb P^{v,a}-a.s.$,
\begin{align*}
V_t^{A,w}=&-U_A(\xi_T)-\int_t^TR_AV_s^{A,w}\left(\mu_s(v_s^{\mathbb P^{v}},a_s^{\mathbb P^{v,a}})\cdot  Z_s^{\obs}+\ell_s(v_s^{\mathbb P^{v}},a_s^{\mathbb P^{v,a}})\cdot  Z_s^{\nobs}-k(v_s^{\mathbb P^{v}},a_s^{\mathbb P^{v,a}})\right)ds\\
&+\int_t^TR_AV_s^{A,w} Z^{\obs}_s\cdot\Sigma_s(v_s^{\mathbb P^{v}})dW_s+\int_t^TR_AV_s^{A,w} Z^{\nobs}_s\cdot\Sigma_s^\perp(v_s^{\mathbb P^{v}})dW_s -\int_t^TR_AV_s^{A,w}dA_s^{v,a}.
\end{align*}
Under this form, we see that the triplet $(V^{A,w},(Z^\obs,Z^{\nobs}), (\overline{A}^{v,a})_{(v,a)\in\mathcal U})$, with
$$\overline A^{v,a}_t:=-R_A\int_0^tV_s^{A,w}dA_s^{v,a},$$
is a solution to the (linear) second-order backward stochastic differential equation (2BSDE for short), as introduced by Soner, Touzi \& Zhang (2012), with terminal condition $U_A(\xi_T)$ and generator $F:[0,T]\times \mathbb R\times \mathbb R^{d_0+1}\times \mathbb R^{d-d_0-1},\times \mathcal V\times\mathcal A\longrightarrow \mathbb R$, defined by
$$F(t,y,z^\obs,z^\nobs,v,a):=R_Ay\left(\mu_t(v,a)\cdot z^\obs+\ell_t(v,a)\cdot z^\nobs- k(v,a)\right).$$
Indeed, according to Definition 3.1 in Soner, Touzi \& Zhang (2012), the only thing that we have to check is that the family of non-decreasing processes $ (\overline{A}^{v,a})_{(v,a)\in\mathcal U})$ satisfies the so-called {\it minimality} condition
$$\overline{A}_t^{v,a}=\underset{(v',a')\in\mathcal U(t,(v,a))}{{\rm essinf}^{\mathbb P^{v,a}}}\ \mathbb E_t^{\mathbb P^{v',a'}}\left[\overline{A}_T^{v',a'}\right],\ \text{for any $(v,a)\in\mathcal U$.}$$
However, this property can be immediately deduced from the definition of the value function $V^{A,w}$ as an $\rm essup$ (see Step (ii) of the proof of Theorem 4.6 in Soner, Touzi \& Zhang (2012) for similar arguments). Moreover, solving the above  equation, we get
\begin{align}\label{eq:value function}
\nonumber V_t^{A,w}=&\ V_0^{A,w}\exp\left[-R_A\int_0^t\left( k(v_s^{\mathbb P^{v}},a_s^{\mathbb P^{v,a}})-\mu_s(v^{\mathbb P^{v}}_s,a^{\mathbb P^{v,a}}_s)\cdot Z_s^\obs-\ell_s(v^{\mathbb P^{v}}_s,a^{\mathbb P^{v,a}}_s)\cdot Z_s^\nobs\right)ds\right]\\
\nonumber&\times\exp\left[-\frac{R_A^2}{2}\int_0^t\left(\No{\Sigma_s^T(v_s^{\mathbb P^{v}})Z^\obs_s}^2+\No{Z_s^\nobs}^2\right)ds+R_AA_t^{v,a}\right]\\
&\times\exp\left[-R_A\left(\int_0^tZ_s^\obs\cdot\Sigma_s(v_s^{\mathbb P^{v}})dW_s+\int_0^tZ_s^\nobs\cdot\Sigma_s^\perp(v_s^{\mathbb P^{v}})dW_s\right)\right],\ \mathbb P^{v,a}-a.s.
\end{align}
The difficulty is that, a priori, we do not know anything about the non-decreasing processes $A^{v,a}$, and thus, it is not in general possible to characterize further the optimal choice of the agent. We will show below that this difficulty disappears when the volatility is not controlled. When it is controlled, there  are still cases for which  it is possible to obtain further information about $A^{v,a}$, which correspond basically to the situations where the contract $\xi_T$ is smooth enough. So far, the most general result in this direction has been obtained by Peng, Song and Zhang (2014), and can be written in our context as the following assumption,  denoting,
\begin{align*}
G^\obs_{d_0}(s,z,\gamma)&:=\underset{(v,a)\in\mathcal V\times\mathcal A}{\sup}g_0^\obs(s,z,\gamma,v,a)\\
&:=\underset{(v,a)\in\mathcal V\times\mathcal A}{\sup}\left\{\frac12{\rm Tr}\left[\gamma\Sigma_s(v)\Sigma_s^T(v)\right]+\mu_s(v,a)\cdot z-k(v,a)\right\}.
\end{align*}

\begin{Assumption}\label{assump:reg}
The contract $\xi_T$ is such that there exists a $\mathbb F^\obs$-predictable process $\Gamma^\obs$ taking values in the set of real $(d_0+1)\times (d_0+1)$ matrices, such that the non-decreasing process $A^{v,a}$ in the decomposition \eqref{eq:value function} is of the form
\begin{align*}
A_t^{v,a}=& \int_0^t\left(
G^\obs_{d_0}(s,Z^\obs_s,\Gamma^\obs_s)-g_{d_0}^\obs(s,Z^\obs_s,\Gamma_s^\obs,v_s^{\mathbb P^{v}},a_s^{\mathbb P^{v,a}})\right)ds,\ \mathbb P^{v,a}-a.s.
\end{align*}
\end{Assumption}
\noindent We now recall that  $\xi_T$ is assumed to be $\mathbb F^\obs$-measurable. Hence, under Assumption \ref{assump:reg}, plugging the explicit expression of $A^{v,a}$ into \eqref{eq:value function}, we see that we necessarily have $Z^\nobs=0$ and $\Gamma^\nobs=0$. We deduce from the representation of $A^{v,a}$ and \eqref{eq:value function} that for any $\mathbb P\in\mathcal P$
\begin{align*}
U_A(\xi_T)=&\ V_0^{A,w}\exp\left[-R_A\int_0^T\left(\frac12{\rm Tr}\left[\left(\Gamma_s^\obs+R_AZ_s^\obs(Z_s^\obs)^T\right)d\langle B^\obs\rangle_s\right]-G_{d_0}^\obs(s,Z_s^\obs,\Gamma_s^\obs)\right)ds\right]\\
&\times\exp\left[-R_A\int_0^TZ_s^\obs\cdot dB^\obs_s\right],\ \mathbb P^{v,a}-a.s.
\end{align*}
Now, let us define for any $(v,a)\in\mathcal U$
$$Z^{\obs,v,a}(B_\cdot):=Z^\obs (B^{\obs,v,a}(B_\cdot)),\ \Gamma^{\obs,v,a}(B_\cdot):=\Gamma^\obs(B^{\obs,v,a}(B_\cdot)).
$$
Then, we deduce by definition of $\mathbb P^{v,a}$ that
\begin{align*}
U_A(\xi_T)=&\ -V_0^{A,w}\exp\left[-R_A\int_0^T\frac12{\rm Tr}\left[\left(\Gamma_s^{\obs,v,a}+R_AZ_s^{\obs,v,a}(Z_s^{\obs,v,a})^T\right)d\langle B^{\obs,v,a}\rangle_s\right]ds\right]\\
&\times\exp\left[-R_A\left(-\int_0^TG_{d_0}^\obs(s,Z_s^{\obs,v,a},\Gamma_s^{\obs,v,a})ds+\int_0^TZ_s^{\obs,v,a}\cdot dB^{\obs,v,a}_s\right)\right],\ \mathbb P_0-a.s.,
\end{align*}
which is exactly the form given in \eqref{form2}.

\vspace{0.5em}
\noindent {\bf Step 3: The case of uncontrolled volatility (Holmstrom-Milgrom 1987).}

\vspace{0.5em}
\noindent Let us assume now that the set $\mathcal V$ is reduced to the singleton $\{v_0\}\subset\mathbb R^m$. In this case, the value function of the agent can be rewritten, for any admissible $a$, as
\begin{equation}\label{eq:valueweak2}
V^{A,w,a}_t:=\underset{(v_0,a^{'})\in\mathcal U(t,(v_0,a))}{{\rm essup}}\ \mathbb E_t^{\mathbb P^{v_0,a'}}\left[U_A(\xi_T)e^{R_AK_{t,T}^{v_0,a'}}\right],\ \mathbb P^{v_0}-a.s.
\end{equation}

\vspace{0.5em}
\noindent For any $a'$ such that $(v_0,a')\in\mathcal U(t,(v_0,a))$, define now
$$V^{A,w,a,a'}_t:= \mathbb E_t^{\mathbb P^{v_0,a'}}\left[U_A(\xi_T)e^{R_AK_{t,T}^{v_0,a'}}\right],\ \mathbb P^{v_0}-a.s.$$
Then, $V^{A,w,a,a'}_t$ is an $\mathbb F$-martingale under $\mathbb P^{v_0,a'}$, so that by the martingale representation, there are $\mathbb R^{d_0}$ and $\mathbb R^{d-d_0-1}$-valued predictable processes $\tilde Z^{\obs,a,a'}$ and $\tilde Z^{\nobs,a,a'}$ such that
\begin{align*}
V_t^{A,w,a,a'}=&\ U_A(\xi_T)+\int_t^TR_AV_s^{A,w,a,a'} k(v_0,a'_s)ds-\int_t^Te^{-R_AK_{0,s}^{v_0,a'}}\tilde Z^{\obs,a,a'}_s\cdot\Sigma_s(v_0)dW^{a'}_s\\
&-\int_t^Te^{-R_AK_{0,s}^{v_0,a'}}\tilde Z^{\nobs,a,a'}_s\cdot\Sigma_s^\perp(v_0)dW^{a'}_s,\ \mathbb P^{v_0}-a.s.
\end{align*}
By definition of $W^{a'}$, we deduce that, $\mathbb P^{v_0}-a.s.,$
\begin{align*}
V_t^{A,w,a,a'}=&\ U_A(\xi_T)-\int_t^TR_AV_s^{A,w,a,a'}\Big(\mu_s(v_0,a'_s)\cdot  Z_s^{\obs,a,a'}+\ell_s(v_0,a'_s)\cdot  Z_s^{\nobs,a,a'}-k(v_0,a'_s)\Big)ds\\
&+\int_t^TR_AV_s^{A,w,a,a'} Z^{\obs,a,a'}_s\cdot\Sigma_s(v_0)dW_s+\int_t^TR_AV_s^{A,w,a,a'} Z^{\nobs,a,a'}_s\cdot\Sigma_s^\perp(v_0)dW_s,
\end{align*}
where we defined
$$Z^{\obs,a,a'}_t:=-\frac{e^{-R_AK_{0,t}^{v_0,a'}}}{R_AV_t^{A,w,a,a'}}\tilde Z^{\obs,a,a'}_t,\ Z^{\nobs,a,a'}_t:=-\frac{e^{-R_AK_{0,t}^{v_0,a'}}}{R_AV_t^{A,w,a,a'}}\tilde Z^{\nobs,a,a'}_t.$$
Let us now simplify everything a bit by setting (remember that $V_t^{A,w,a,a'}$ is negative by definition)
$$Y_t^{a,a'}:=-\frac{\ln\left(-V_t^{A,w,a,a'}\right)}{R_A}.$$
Then by It\^o's formula, we deduce that the following holds, $\mathbb P^{v_0}-a.s.,$
\begin{align*}
Y_t^{a,a'}=&\ \frac{\log(-U_A(\xi_T))}{R_A}+\int_t^T\Big(-\frac{R_A}{2}\No{\Sigma_s^T(v_0)Z_s^{\obs,a,a'}}^2+\mu_s(v_0,a'_s)\cdot Z_s^{\obs,a,a'}+\ell_s(v_0,a'_s)\cdot Z_s^{\nobs,a,a'}\\
&-k(v_0,a'_s)\Big)ds-\int_t^TZ_s^{\obs,a,a'}\cdot\Sigma_s(v_0)dW_s-\int_t^TZ_s^{\nobs,a,a'}\cdot\Sigma_s^\perp(v_0)dW_s.
\end{align*}
The above equation can be identified as a linear-quadratic backward SDE with terminal condition $\log(-U_A(\xi_T))/R_A$. Therefore, using the theory of BSDEs with quadratic growth, and in particular the corresponding comparison theorem (see for instance El Karoui, Peng, Quenez (1997), Kobylanski (2000) and Briand and Hu (2008)), we deduce that if $U_A(\xi_T)$ has second order moments under $\mathbb P^{v_0}$ and if we define
$$Y_t^a:=\underset{(v_0,a')\in\mathcal U(t,(v_0,a))}{\rm essup}\ Y_t^{a,a'},$$
then there exists a $\mathbb F^\obs$-predictable process $Z^a$ and a $\mathbb F^\nobs$-predictable process $Z^{\nobs,a}$ such that, $\mathbb P^{v_0}-a.s.$,
\begin{align*}
Y_t^{a}=&\ \frac{\log(-U_A(\xi_T))}{R_A}+\int_t^T\underset{a\in\mathcal A}{\sup}\left\{\mu_s(v_0,a)\cdot Z_s^{\obs,a}+\ell_s(v_0,a)\cdot Z^{\nobs,a}_s-k(v_0,a)\right\}ds\\
&-\int_t^T\frac{R_A}{2}\left(\No{\Sigma_s^T(v_0)Z_s^{\obs,a}}^2+\No{Z_s^{\nobs,a}}^2\right)ds-\int_t^T Z^{\obs,a}_s\cdot\Sigma_s(v_0)dW_s-Z^{\nobs,a}_s\cdot\Sigma_s^\perp(v_0)dW_s,
\end{align*}
that is $(Y^a,Z^{\obs,a},Z^{\nobs,a})$ solves the backward SDE with terminal condition $\log(-U_A(\xi_T))/R_A$ and generator $f$, where
$$f(s,z^1,z^1):=-\frac{R_A}{2}\left(\No{\Sigma_s^T(v_0)z^1}^2+\No{z^2}^2\right)+\underset{a\in\mathcal A}{\sup}\left\{\mu_s(v_0,a)\cdot z^1+\ell_s(v_0,a)\cdot z^2-k(v_0,a)\right\}.$$
 By the assumptions of the theorem on then cost function $k$, it can be easily shown, using the linear growth of $\mu$ and $\ell$ in $a$, that the $\sup$ in the definition of $f$ is always attained for some $a^*(s,z^1,z^2)$ that satisfies
$$\No{a^*(s,z^1,z^2)}\leq C_0\left(1+\No{z^1}^{\frac1\varepsilon}+\No{z^2}^{\frac1\varepsilon}\right).$$
In particular this implies that the above BSDE is quadratic in $z$, and is therefore indeed well-posed. Finally, we deduce that
\begin{align*}\nonumber V_t^{A,w,a}=&\ V_0^{A,w}\exp\left[-R_A\int_0^t\left( k(v_0,a_s)-\mu_s(v_0,a_s)\cdot Z_s^{\obs,a}-\ell_s(v_0,a_s)\cdot Z_s^{\nobs,a}\right)ds\right]\\
\nonumber&\times\exp\left[-\frac{R_A^2}{2}\int_0^t\left(\No{\Sigma_s^T(v_0)Z^{\obs,a}}^2+\No{Z_s^{\nobs,a}}^2\right)ds+R_AA_t^{a}\right]\\
&\times\exp\left[-R_A\left(\int_0^tZ_s^{\obs,a}\cdot\Sigma_s(v_0)dW_s+\int_0^tZ_s^{\nobs,a}\cdot\Sigma_s^\perp(v_0)dW_s\right)\right],\ \mathbb P^{v_0}-a.s.,
\end{align*}
where the non-decreasing process $A^a$ is defined by
\begin{align*}
A_t^{a}=& \int_0^t\underset{a\in\mathcal A}{\sup}\Big\{\mu_s(v_0,a)\cdot Z_s^{\obs,a}+\ell_s(v_0,a)\cdot Z_s^{\nobs,a}-k(v_0,a)\\
&\hspace{3.5em}-\frac{R_A}{2}\left(\No{\Sigma_s^T(v_0)Z^{\obs,a}_s}^2+\No{Z_s^{\nobs,a}}^2\right)\Big\}ds\\
&-\int_0^t\Big(\mu_s(v_0,a_s)\cdot Z_s^{\obs,a}+\ell_s(v_0,a_s)\cdot Z_s^{\nobs,a}-k(v_0,a_s)\\
&\hspace{2em}-\frac{R_A}{2}\left(\No{\Sigma_s^T(v_0)Z^{\obs,a}_s}^2+\No{Z_s^{\nobs,a}}^2\right)\Big)ds,\ \mathbb P^{v_0}-a.s.
\end{align*}
Again, we must have $Z^{\nobs,a}=0$ in order to ensure that $\xi_T$ is $\mathcal F_T^\obs$-measurable, so that we have
$$A_t^{a}= \int_0^t\left(
G^\obs_{d_0}(s,Z^\obs_s,0)-g_{d_0}^\obs(s,Z^\obs_s,0,v_0,a_s)\right)ds,\ \mathbb P^{v_0}-a.s.,$$
which means that Assumption \ref{assump:reg} is satisfied.
\ep

\vspace{0.5em}
\noindent {\bf Proof of Proposition \ref{prop.growth}.} Notice first that the assumption on $k$ implies that it has a super quadratic growth at infinity, which means that for every $(z,\gamma)\in\mathbb R^2\times\mathbb S^2$, the infimum in the definition of $G_1^\obs(z,\gamma)$ is always attained for at least one $v^*(z,\gamma)$. Moreover, as it is then an interior maximizer, it necessarily satisfies the first-order conditions, which can be rewritten here as
\begin{equation}\label{eq.growth}
Mz+\gamma \tilde Mv^*(z,\gamma)-\nabla k(v^*(z,\gamma))=0,
\end{equation}
for some matrices $M$ and $\tilde M$, independent of $(z,\gamma)$.

\vspace{0.5em}
\noindent In particular, the above shows that $v^*(z,\gamma)$ cannot remain bounded as $\|z\|$ and $\|\gamma\|$ go to $+\infty$. Let us also verify that we have
$$\|v^*(z,\gamma)\|\leq C_0\left(1+\|z\|^{\frac{1}{1+\varepsilon}}+\|\gamma\|^{\frac{1}{\varepsilon}}\right),$$
for some constant $C_0>0$.

\vspace{0.5em}
\noindent Indeed assume first that $\|v^*(z,\gamma)\|/\|z\|^{\frac{1}{1+\varepsilon}}$ does not remain bounded when $\|z\|$ goes to $+\infty$. Then, we deduce from \eqref{eq.growth} that
$$M\frac{z}{\|z\|^{\frac{1}{1+\varepsilon}}}+\gamma \tilde M\frac{v^*(z,\gamma)}{\|z\|^{\frac{1}{1+\varepsilon}}}-\frac{\nabla k(v^*(z,\gamma))}{\|v^*(z,\gamma)\|^{1+\varepsilon}}\frac{\|v^*(z,\gamma)\|^{1+\varepsilon}}{\|z\|}\|z\|^{\frac{\varepsilon}{1+\varepsilon}}=0.$$
As $\|z\|$ goes to $+\infty$, the third term above then clearly dominates the other two, which contradicts the fact that their sum should be $0$.

\vspace{0.5em}
\noindent Similarly, if we assume that $\|v^*(z,\gamma)\|/\|\gamma\|^{\frac{1}{\varepsilon}}$ does not remain bounded when $\|\gamma\|$ goes to $+\infty$, we deduce that
$$M\frac{z}{\abs{\gamma}^{\frac{1}{\varepsilon}}}+\gamma \tilde M\frac{v^*(z,\gamma)}{\|\gamma\|^{\frac{1}{\varepsilon}}}-\frac{\nabla k(v^*(z,\gamma))}{\|v^*(z,\gamma)\|^{1+\varepsilon}}\frac{\|v^*(z,\gamma)\|^{1+\varepsilon}}{\|\gamma\|^{\frac{1+\varepsilon}{\varepsilon}}}\|\gamma\|=0.$$
Again, the third term dominates the others as $\|\gamma\|$ goes to $+\infty$, which contradicts the equality.

\vspace{0.5em}
\noindent We next deduce that for every $\eta>0$
$$\frac{\|v^*(z,\gamma)\|}{1+\|z\|^{\frac{1}{1+\varepsilon}-\eta}+\|\gamma\|^{\frac{1}{\varepsilon}-\eta}}\ \text{is not bounded as $\|z\|$ and $\|\gamma\|$ go to $+\infty$.}$$
Indeed, if we assume that $\|v^*(z,\gamma)\|/\|z\|^{\frac{1}{1+\varepsilon}-\eta}$ remains bounded near infinity, then dividing \eqref{eq.growth} by $\|z\|^{1/(1+\varepsilon)-\eta}$, we obtain that the first term behaves, as $\|z\|$ goes to $+\infty$, like $\|z\|^{\varepsilon/(1+\varepsilon)+\eta}$, while the second one is bounded and the third one behaves like  $\|z\|^{\varepsilon/(1+\varepsilon)-\varepsilon\eta}$. Hence, the first term dominates and we again have a contradiction. The result for the growth with respect to $\gamma$ is proved in the same manner.

\vspace{0.5em}
\noindent From the above growth for $v^*(z,\gamma)$, it is clear that the dominating terms at infinity in \eqref{finalopt2} are
$$\frac{1}{2}\No{  \theta^*(Z,\Gamma)}^2R_A\left(Z^X\right)^2,\  \frac{R_P}{2}\No{\theta^*(Z,\Gamma)}^2\left(1-Z^X\right)^2,\ \text{and}\ k(v^*(z,\gamma)),$$
which are all non-negative. In particular, \eqref{finalopt2} goes to $+\infty$ as $\|z\|$ and $\|\gamma\|$ go to $+\infty$, and the minimum is therefore attained.
\ep


\vspace{0.5em}

\begin{Proposition}\label{prop:max} {\bf (Sufficient conditions for the existence of optimal $v$ in the  non-contractible case.)}
Suppose $a=0$ and $B^1$  cannot be contracted upon.
Let $\mathcal I$ be the subset of $\{1,\dots,d\}$ such that for every $j\in\mathcal I$, $\beta_j=\min_i\beta_i$ and assume
that $b_j\neq 0$ for  $j\in\mathcal I$. If either of the following holds

\vspace{0.5em}
(i) ${\rm Card}(\mathcal I)>1$ and there is at least one pair $(i,j)\in\mathcal I\times\mathcal I$ such that $\alpha_{j}\beta_{j}/b_{j}\neq \alpha_i\beta_i/b_i$.

\vspace{0.5em}
(ii) ${\rm Card}(\mathcal I)\geq 1$, for all $(i,j)\in\mathcal I\times\mathcal I$, $i\neq j$, $\alpha_j\beta_j/b_j=\alpha_i\beta_i/b_i=:\eta$ and
\begin{align*}
&\frac{(1+\eta)^2\left(\sum_{i\in\mathcal I}b_i\right)^2}{2\left({\rm Card}(\mathcal I)R_A\eta^2+{\rm Card}(\mathcal I)R_P\left(1-\eta\right)^2+\sum_{i\in\mathcal I}\beta_i\right)}+\sum_{i\notin \mathcal I}(b_i+\alpha_i\beta_i)\frac{-\eta b_i+\alpha_i\beta_i}{\beta_i-\min_j\beta_j}\\
&-\frac12\sum_{i\notin\mathcal I}\frac{(-\eta b_i+\alpha_i\beta_i)^2}{(\beta_i-\min_j\beta_j)^2}\left(R_A\eta^2+R_P\left(1-\eta\right)^2+\beta_i\right)\leq \sum_{i=1}^d\frac{(b_i+\alpha_i\beta_i)^2}{2\left(\beta_i+R_A\right)}.
\end{align*}
Then, there exists at least one couple $(Z^*,\Gamma^*)\in\mathbb R\times(-\infty,\min_i\beta_i)$  attaining the maximum in the principal's problem \footnote{Notice that the left-hand side in (ii) above can be made negative, if for instance ${\rm Card}(\mathcal I)\leq d-1$ and $\min_{i\notin\mathcal I}\beta_i$ is sufficiently close to $\min_i\beta_i$, so that the  condition can, indeed, be satisfied in examples.}
.
\end{Proposition}

\noindent {\bf Proof of Proposition \ref{prop:max}.} If for some $i$, we have $\Gamma -  \beta_i >0 $, or $\Gamma -  \beta_i =0 $, but $Z\neq0$, it is easily verified that the agent chooses optimally $|  v _i^*(Z,\Gamma)|=\infty$,
and that this cannot be optimal for the principal. Thus,  we can  optimize under the constraint
$\Gamma -  \min_j\beta_j \le 0  $.

\vspace{0.5em}
\noindent In the case in which  $\Gamma-  \beta_i = 0 $ for some $i$ and $Z=0$,
 the principal has to
maximize over all admissible values of $v_i$, because the agent is indifferent among those.
It can be verified that for $Z=0$ and in that case there exists an optimal $v^*$ for the problem \eqref{finaloptB}.

\vspace{0.5em}
\noindent
The only remaining case is
  $\Gamma
  -\beta_i
   < 0$ for all $i$.
   Then,
   the contract is incentive compatible for
$$v^*_i(Z,\Gamma)=\frac{b_iZ+\alpha_i\beta_i}{
\beta_i -
\Gamma}~,~~i=1, \ldots, d.$$
From \reff{finaloptB}, the principal's problem is then to maximize
$$v^*\cdot b-\frac12\No{v^*}^2\left(R_AZ^2+R_P(1-Z)^2\right)-\frac12\sum_{i=1}^d\beta_i\left(v^*_i-\alpha_i\right)^2,$$
which is the same as maximizing
\begin{equation}\label{eq:secondbestB}
f(Z,\Gamma):=\sum_{i=1}^d(b_i+\alpha_i\beta_i)\frac{b_iZ+\alpha_i\beta_i}{\beta_i-\Gamma}-\frac12\sum_{i=1}^d\frac{(b_iZ+\alpha_i\beta_i)^2}{(\beta_i-\Gamma)^2}\left(R_AZ^2+R_P(1-Z)^2+\beta_i\right).
\end{equation}
We will show that the maximization of \eqref{eq:secondbestB} over all $(Z,\Gamma)\in\mathbb R\times (-\infty,\min_i\beta_i)$ can be reduced to a maximization over a compact set strictly included in $\mathbb R\times(-\infty,\min_i\beta_i)$. First, notice that by taking $(Z,\Gamma)=(1,-R_A)$, \reff{eq:secondbestB} becomes
$$\sum_{i=1}^d\frac{(b_i+\alpha_i\beta_i)^2}{2\left(\beta_i+R_A\right)}\geq0,$$
which implies that the maximum in \reff{eq:secondbestB} is non-negative.

\vspace{0.5em}
\noindent Let us now show that the maximum in \reff{eq:secondbestB} can never be achieved on the boundary of the domain $\mathbb R\times(-\infty,\min_i\beta_i)$. Let us distinguish several cases:

\vspace{0.5em}
\noindent $(i)$ $Z$ goes to $\pm\infty$ and $\Gamma$ remains bounded and does not go to $\min_i\beta_i$. Then, it is easily seen that $f$ goes to $-\infty$. This is therefore suboptimal.

\vspace{0.5em}
\noindent $(ii)$ $Z$ remains bounded and $\Gamma$ goes to $-\infty$. Then, $f$ goes to $0$ which is again suboptimal.

\vspace{0.5em}
\noindent $(iii)$ $Z$ goes to $\pm\infty$ and $\Gamma$ goes to $-\infty$. Then, $f$ can either go to $-\infty$ or $0$, depending on whether $Z/\Gamma$ remains bounded or not.

\vspace{0.5em}
\noindent $(iv)$ $\Gamma$ goes to $\min_i\beta_i$ and $Z$ goes to $\pm\infty$. Then, $f$ goes to $-\infty$.

\vspace{0.5em}
\noindent $(v)$ $\Gamma$ goes to $\min_i\beta_i$ and $Z$ remains bounded. Then, we have to distinguish between three sub-cases.
\begin{itemize}
\item[$\bullet$] If $Z$ is fixed and $Z\neq -\alpha_j\beta_j/b_j$, for every $j\in\mathcal I$, then, $f$ goes to $-\infty$.
\item[$\bullet$] If $Z$ goes to $-\alpha_j\beta_j/b_j$ for some $j\in\mathcal I$, ${\rm Card}(\mathcal I)>1$ and there is at least one $j_0\in\mathcal I\backslash\{j\}$ such that $\alpha_{j_0}\beta_{j_0}/b_{j_0}\neq \alpha_j\beta_j/b_j$. Then, $f$ still goes to $-\infty$.
\item[$\bullet$] If $Z$ goes to $-\alpha_j\beta_j/b_j$ for some $j\in\mathcal I$, ${\rm Card}(\mathcal I)\geq 1$ and for all $(i,j)\in\mathcal I\times\mathcal I$, $i\neq j$, $\alpha_j\beta_j/b_j=\alpha_i\beta_i/b_i=:\eta$,
then, if $\frac{b_jZ+\alpha_j\beta_j}{\beta_j-\Gamma}$ goes to $\pm\infty$, then, $f$ goes to $-\infty$. If $\frac{b_jZ+\alpha_j\beta_j}{\beta_j-\Gamma}$ remains bounded, then, the maximum value that can be achieved by $f$ is
\begin{align*}
&\sup_{u\in\mathbb R}\left\{(1+\eta)\sum_{i\in\mathcal I}b_iu-\frac12u^2\left({\rm Card}(\mathcal I)R_A\eta^2+{\rm Card}(\mathcal I)R_P\left(1-\eta\right)^2+\sum_{i\in\mathcal I}\beta_i\right)\right\}\\
&+\sum_{i\notin \mathcal I}^d(b_i+\alpha_i\beta_i)\frac{-\frac{\alpha_j\beta_jb_i}{b_j}+\alpha_i\beta_i}{\beta_i-\beta_j}\\
&-\frac12\sum_{i\notin \mathcal I}^d\frac{(-\frac{\alpha_j\beta_jb_i}{b_j}+\alpha_i\beta_i)^2}{(\beta_i-\beta_j)^2}\left(R_A\left(\frac{\alpha_j\beta_j}{b_j}\right)^2+R_P\left(1-\frac{\alpha_j\beta_j}{b_j}\right)^2+\beta_i\right),
\end{align*}
which is equal to
\begin{align}\label{fmax}
\nonumber &\frac{(1+\eta)^2\left(\sum_{i\in\mathcal I}b_i\right)^2}{2\left({\rm Card}(\mathcal I)R_A\eta^2+{\rm Card}(\mathcal I)R_P\left(1-\eta\right)^2+\sum_{i\in\mathcal I}\beta_i\right)}+\sum_{i\notin \mathcal I}(b_i+\alpha_i\beta_i)\frac{-\eta b_i+\alpha_i\beta_i}{\beta_i-\min_j\beta_j}\\
&-\frac12\sum_{i\notin\mathcal I}\frac{(-\eta b_i+\alpha_i\beta_i)^2}{(\beta_i-\min_j\beta_j)^2}\left(R_A\eta^2+R_P\left(1-\eta\right)^2+\beta_i\right).
\end{align}
However, by assumption, this is again sub-optimal.
\end{itemize}
\ep


\begin{thebibliography}{cea}
\bibitem{bichteler}
Bichteler, K. (1981). Stochastic integration and $L^p$-theory of semimartingales, {\sl Ann. Prob.}, 9(1):49--89.

\bibitem{bh}
Briand, P., Hu, Y. (2008). Quadratic BSDEs with convex generators and unbounded terminal conditions, {\sl Prob. Theory and Related Fields}, 141(3-4):543--567.

\bibitem{CCZ}Cadenillas, A., Cvitani\'c, J., Zapatero, F. (2007). Optimal risk-sharing
with effort and project choice, {\it Journal of Economic Theory}, 133:403--440.

\bibitem{cf}
Cont, R., Fourni\'e, D. (2013). Functional It\^o calculus and stochastic integral representation of martingales, {\sl Annals of Probability}, 41(1):109--133.

\bibitem{CWY} Cvitani\'c, J.,   Wan, X., Yang, H. (2013). Dynamics of contract design with screening, {\it Management Science,} 59:1229--1244.

\bibitem{Dupire}
Dupire, B. (2009). Functional It\^o calculus, preprint.

\bibitem{epq}
El Karoui, N., Peng, S., Quenez, M.-C. (1997). Backward stochastic differential equations in finance, {\sl Mathematical Finance}, 7:1--71.

\bibitem{eq}
El Karoui, N., Quenez, M.-C. (1995). Dynamic programming and pricing of contingent claims in an incomplete market, {\sl SIAM J. Control Optim.}, 33(1):29--66.

\bibitem{ej}
Epstein, L., Ji, S. (2013). Ambiguous volatility and asset pricing in continuous time, {\sl Rev. Financ. Stud.}, 26:1740--1786.

\bibitem{ej2}
Epstein, L., Ji, S. (2014). Ambiguous volatility, possibility and utility in continuous time, {\sl Journal of Mathematical Economics},
50:269--282

\bibitem{HM} Holmstrom B., Milgrom, P. (1987). Aggregation and linearity in the
provision of intertemporal incentives, {\em Econometrica}, 55(2):303--328.
\bibitem{kpz}
Kazi-Tani, N., Possama\"i, D., Zhou, C. (2013). Second-order BSDEJ with jumps: formulation and uniqueness, {\sl Ann. of App. Probability}, to appear.


\bibitem{L} Leung, R.C.W. (2014). Continuous-time principal-agent problem with drift and
stochastic volatility control: with applications to delegated
portfolio management, working paper, {\sl http://ssrn.com/abstract=2482009}.

\bibitem{LP} Lioui, A., Poncet, P (2013). Optimal benchmarking for active portfolio managers,
{\it European Journal of Operational Research,} 226(2):268--276.
\bibitem{kob}
Kobylansky, M. (2000). Backward stochastic differential equations and partial differential equations with quadratic growth, {\sl Ann. Prob.}, 2:558--602.

\bibitem{nn}
Neufeld, A., Nutz, M. (2014). Measurability of semimartingale characteristics with respect to the probability law, {\sl Stoc. Proc. App.}, 124(11):3819--3845.
\bibitem{nutz}
Nutz, M. (2012). Pathwise construction of stochastic integrals, {\sl Elec. Com. Prob.}, 17(24):1--7.

\bibitem{ns}
Nutz, M., Soner, H.M. (2012). Superhedging and dynamic risk measures under volatility uncertainty, {\sl SIAM Journal on Control and Optimization}, 50(4):2065--2089.
   \bibitem{nv} Nutz, M., van Handel, R. (2013). Constructing sublinear expectations on path space, {\sl Stochastic Processes and their Applications}, 123(8):3100--3121.
   \bibitem{pp} Pag\`es, H., Possama\"i, D. (2014). A mathematical treatment of bank monitoring incentives, {\sl Finance and Stochastics}, 18(1):39-73.
   \bibitem{psz}
   Peng, S., Song, Y., Zhang, J. (2014). A complete representation theorem for $G$-martingales, {\sl Stochastics}, 86:609--631.
   \bibitem{prt}
   Possama\"i, D., Royer, G., Touzi, N. (2013). On the robust super hedging of measurable claims, {\sl  Elec. Comm. Prob.}, 18(95):1--13.

\bibitem{} Ou-Yang, H. (2003). Optimal contracts in a continuous-time
delegated portfolio management problem, {\it The Review of Financial
Studies}, 16:173--208.

\bibitem{Sann} Sannikov, Y. (2008). A continuous-time version of the Principal-Agent
problem, {\it Review of Economic Studies}, 75:957--984.

      \bibitem{stz3} Soner, H.M., Touzi, N., Zhang, J. (2011). Quasi-sure stochastic analysis through aggregation, {\sl Electronic Journal of Probability}, 16(67):1844--1879.
       \bibitem{stz2} Soner, H.M., Touzi, N., Zhang, J. (2013). Dual formulation of second order target problems, {\sl Annals of Applied Probability}, 23(1):308--347.
       \bibitem{stz1}
Soner, H.M., Touzi, N., Zhang, J. (2012) Wellposedness of second-order backward SDEs, {\sl Prob. Th. and Rel. Fields}, 153(1-2):149--190.
\bibitem{Su}
Sung, J. (1995). Linearity with project selection and controllable diffusion rate
in con-tinuous-time principal-agent problems, {\sl RAND J. Econ.} 26, 720--743.

\bibitem{W} Wong, T.-Y. (2013). Dynamic agency and exogenous risk taking, working paper.
\end{thebibliography}
\end{document}